\documentclass[reprint,amsmath,amssymb,pra,superscriptaddress,floatfix]{revtex4-1}

\usepackage{color}
\usepackage{graphicx}
\usepackage{bm}
\usepackage{amsmath,amssymb}
\usepackage{xspace}
\usepackage{braket}
\usepackage{float}
\usepackage{lineno}

\newdimen\linenumbersep
\usepackage{hyperref}

\newcommand*{\SFigref}[1]{\hyperref[#1]{\ref*{#1}}}

\AtBeginDocument{\let\oldcontentsline\contentsline}
\newcommand{\notoccontentsline}[4]{\oldcontentsline{}{}{}{}}
\newcommand{\droptocpage}{\addtocontents{toc}{\let\protect\contentsline\protect\notoccontentsline}}
\newcommand{\incltocpage}{\addtocontents{toc}{\let\protect\contentsline\protect\oldcontentsline}}

\newcommand{\internallinenumbersJ}{}
\newcommand{\linenumbersJ}{}

\begin{document}
    % \runningpagewiselinenumbers
    \linenumbersJ
	\title{Chiral Cavity Quantum Electrodynamics}

	\author{John Clai Owens}
	\affiliation{James Franck Institute and Department of Physics, University of Chicago, Chicago, IL 60637, USA}
	\affiliation{Thomas J. Watson, Sr., Laboratory of Applied Physics and Kavli Nanoscience Institute, California Institute of Technology, Pasadena, California 91125, USA}
	\author{Margaret G. Panetta}
	\affiliation{James Franck Institute and Department of Physics, University of Chicago, Chicago, IL 60637, USA}
	\author{Brendan Saxberg}
	\affiliation{James Franck Institute and Department of Physics, University of Chicago, Chicago, IL 60637, USA}
	\author{Gabrielle Roberts}
	\affiliation{James Franck Institute and Department of Physics, University of Chicago, Chicago, IL 60637, USA}
	\author{Srivatsan Chakram}
	\affiliation{Department of Physics and Astronomy, Rutgers University, Piscataway, NJ 08854, USA}
	\author{Ruichao Ma}
	\affiliation{Department of Physics and Astronomy, Purdue University, West Lafayette, IN 47907, USA}
	\author{Andrei Vrajitoarea}
	\affiliation{James Franck Institute and Department of Physics, University of Chicago, Chicago, IL 60637, USA}
	\author{Jonathan Simon}
	\affiliation{James Franck Institute and Department of Physics, University of Chicago, Chicago, IL 60637, USA}
	\author{David Schuster}
	\affiliation{James Franck Institute and Department of Physics, University of Chicago, Chicago, IL 60637, USA}
	 
\date{\today}

\begin{abstract}

Cavity quantum electrodynamics, which explores the granularity of light by coupling a resonator to a nonlinear emitter~\cite{walther2006cavity}, has played a foundational role in the development of modern quantum information science and technology. In parallel, the field of condensed matter physics has been revolutionized by the discovery of underlying topological robustness in the face of disorder~\cite{von1986quantized,stormer1999fractional,Hasan2010}, often arising from the breaking of time-reversal symmetry, as in the case of the quantum Hall effect. In this work, we explore for the first time cavity quantum electrodynamics of a transmon qubit in the topological vacuum of a Harper-Hofstadter topological lattice~\cite{hofstadter1976energy}. To achieve this, we assemble a square lattice of niobium superconducting resonators~\cite{Reagor2013-highQ} and break time-reversal symmetry by introducing ferrimagnets~\cite{Owens2018} before coupling the system to a single transmon qubit. We spectroscopically resolve the individual bulk and edge modes of this lattice, detect vacuum-stimulated Rabi oscillations between the excited transmon and each mode, and thereby measure the synthetic-vacuum-induced Lamb shift of the transmon. Finally, we demonstrate the ability to employ the transmon to count individual photons~\cite{schuster2007resolving} within each mode of the topological band structure. This work opens the field of chiral quantum optics experiment~\cite{lodahl2017chiral}, suggesting new routes to topological many-body physics~\cite{anderson2016engineering,de2020light} and offering unique approaches to backscatter-resilient quantum communication.

\end{abstract}

\maketitle

\section{Introduction}

Materials made of light are a new frontier in quantum many-body physics~\cite{carusotto2013quantum}; relying upon non-linear emitters to generate strong photon-photon interactions and ultra-low-loss meta-materials to manipulate the properties of the individual photons, this field explores the interface of condensed matter physics and quantum optics whilst simultaneously producing novel devices for manipulating light~\cite{shomroni2014all,baur2014single}. Recent progress in imbuing photons with topological properties~\cite{ozawa2019topological}, wherein the photons undergo circular time-reversal-breaking orbits, promises opportunities to explore photonic analogs of such solid-state phenomena as the (fractional) quantum Hall effect~\cite{von1986quantized,stormer1999fractional}, Abrikosov lattices~\cite{abrikosov1957magnetic}, and topological insulators~\cite{Hasan2010}.

In electronic materials, the circular electron orbits result from magnetic or spin-orbit couplings~\cite{Hasan2010}. Unlike electrons, photons are charge-neutral objects and so do not directly couple to magnetic fields. There is thus an effort to generate synthetic magnetic fields for photons and more generally to explore ideas of topological quantum matter in synthetic photonic platforms. Significant progress in this arena has been made in both optical- and microwave- topological photonics: in silicon photonics~\cite{Rechtsman:2013aa, HafeziM.:2013aa} and optics~\cite{Schine2016,schine2019electromagnetic}, synthetic gauge fields have been achieved while maintaining time-reversal symmetry by encoding a pseudo-spin in either the polarization or spatial mode. In RF and microwave meta-materials, both time-reversal-symmetric~\cite{ningyuan2015time,lu2019probing} and time-reversal-symmetry-broken models have been explored, with the T-breaking induced either by coupling the light to ferrimagnets in magnetic fields~\cite{Wang:2009aa,Owens2018} or by Floquet engineering~\cite{roushan2016chiral}.

To mediate interactions between photons, a nonlinear emitter or ensemble of nonlinear emitters must be introduced into the system~\cite{carusotto2020photonic}. This has been realized for optical photons by coupling them to Rydberg-dressed atoms, providing the first assembly of two-photon Laughlin states of light~\cite{clark2020observation}. In quantum circuits, a 3-site lattice of parametrically-coupled transmon qubits enabled observation of chiral orbits of photons/holes~\cite{roushan2016chiral}, and a $1 \times 8$ lattice of transmons enabled exploration of Mott physics~\cite{ma2019dissipatively}. In nanophotonics, a topological interface enabled helical information transfer between a pair of quantum dots~\cite{barik2018topological}.

	\begin{figure}[!ht]
		%\centering

		\includegraphics[width=.48 \textwidth]{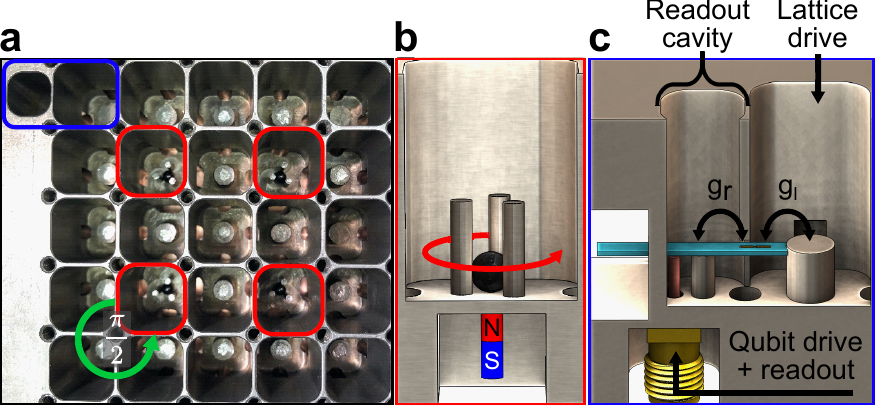} 
		\internallinenumbersJ
		\caption{
		\textbf{Elements of chiral cavity quantum electrodynamics.}
		\textbf{a,} The apparatus consists of a $5\times 5$, $\alpha=\frac{1}{4}$ Hofstadter lattice~\protect{\cite{hofstadter1976energy}} of resonators in which microwave photons propagate as charged particles in a magnetic field, coupled to a single qubit on the edge that is sensitive to the precise number of photons and their energies. Each site, implemented as a coaxial resonator milled into a block of niobium~\protect{\cite{Reagor2013-highQ}}, exhibits a resonance frequency $\omega_0$ determined by the length of a central post, and a nearest neighbor tunneling rate $J$ determined by the size of a machined coupling hole. The synthetic magnetic field manifests as an Aharonov-Bohm flux $\pi/2$ when photons hop around minimal closed loops (green), generated by the spatial structure of the resonator modes: each 2-site by 2-site plaquette includes one lattice site (red) that exhibits a $p_x+i p_y$ orbital, while the other three sites exhibit $s$ orbitals~\protect{\cite{anderson2016engineering,Owens2018}}. The additional site (blue) on the system edge serves as readout cavity into which transmons may be inserted. \textbf{b,} $p_x+i p_y$ sites instead contain \emph{three} posts and thus support three microwave modes ($s$, $p_x\pm i p_y$). Because our Hofstadter lattice employs only the $p_x+i p_y$ mode, we must isolate it: the $s$ mode is tuned away by the electromagnetic coupling between posts, while a Yttrium-Iron-Garnet (YIG) ferrimagnet (black) couples primarily to the $p_x-i p_y$ mode (due to the orientation of the B-field of the red/blue bar magnet), thereby detuning it in energy and isolating the $p_x+i p_y$ mode. \textbf{c,} A transmon qubit is inserted into a gap between readout (left) and lattice (right) cavities on a sapphire carrier (turquoise), and couples to the two cavities with Rabi frequencies $g_r$ and $g_l$ respectively. An SMA connector (gold) allows direct microwave probing of this readout cavity and thus the transmon.}

		\label{fig:LatticeSetup}
	\end{figure}

In this work, we demonstrate a scalable architecture for probing interacting topological physics with light. Building on prior room-temperature work~\cite{Owens2018}, we demonstrate a $5\times 5$ array of superconducting resonators that acts as a quarter flux ($\alpha=\frac{1}{4}$) Hofstadter lattice~\cite{hofstadter1976energy}, exhibiting topological bulk and edge modes for the photons that reside within it. We couple a single transmon qubit to the edge of this system, and enter, for the first time, the regime of strong-coupling cavity quantum electrodynamics for a highly nonlinear emitter interacting with the spectrally resolved modes of a topological band structure.

In Section~\ref{Sec:MagField}, we introduce our superconducting topological lattice architecture compatible with transmon qubits. We then characterize its properties both spectroscopically and spatially in Section~\ref{Sec:ProbeLattice}. In Section~\ref{Sec:Qubit}, we couple a single transmon qubit to the lattice, employing it to detect and manipulate individual photons in bulk and edge modes of the lattice and to measure the Lamb shift of this synthetic vacuum. In Section~\ref{Sec:Conclusion} we conclude, exploring the opportunities opened by this platform.

\section{A Superconducting Hofstadter Lattice for Microwave Photons}
\label{Sec:MagField}
	
In vacuum, photons are neither (i) massive, nor (ii) charged, nor (iii) confined to two dimensions, the crucial ingredients for quantum Hall physics~\cite{von1986quantized}. To realize these essentials, we follow the road-map laid out in~\cite{anderson2016engineering}: microwave photons are trapped in a 2D array of microwave resonators, and thereby confined to two transverse dimensions and imbued with an effective mass due to the finite tunneling rate between the resonators. Rather than attempting to \emph{actually} imbue photons with electric charge, we note that when electrons are confined to a lattice, the entire impact of a magnetic field on their dynamics is encompassed by the Aharonov-Bohm-like phase that they acquire when tunneling around closed trajectories. We engineer this ``Peierls phase'' via the spatial structure of the on-site lattice orbitals.

Fig.~\ref{fig:LatticeSetup}a shows the square Hofstadter lattice that we have developed for this work. Each square in the diagram is a lattice site, implemented as a resonator of frequency $\omega_0 \approx 2\pi\times 9$ GHz, tunnel-coupled to its nearest-neighbors with $J=2\pi\times 18$ MHz (see SI~\ref{SI:Lattice Fab}). Sites with counter-clockwise red arrows exhibit modes with spatial structure $p_x+i p_y$, while all other sites have $s$-like modes. The phase winding in a $p_x+i p_y$ site causes photons tunneling in/out from different directions to acquire a phase $\phi=\delta\theta$, where $\delta\theta$ is the angle between input and output directions~\cite{Owens2018}. This ensures that when photons tunnel around a closed loop enclosing $n$ plaquettes, they pick up an Aharonov-Bohm-like phase $\phi_{loop}=n\frac{\pi}{2}$. Such a tight-binding model with a flux per plaquette of $\pi/2$ is called a ``quarter flux Hofstadter lattice''~\cite{hofstadter1976energy}.

To avoid seam losses and thus achieve the highest quality factors, the lattice structure is machined from a single block of high-purity (RRR=300) niobium and cooled to $30$ mK to reduce loss and eliminate blackbody photons (see SI~\ref{SI:DilFridgeWiring}). Sites are realized as coaxial resonators, while tunneling between adjacent sites is achieved via hole-coupling through the back side: as in~\cite{chakram2020seamless}, the coupling holes are sub-wavelength and thus do not lead to leakage out of the structure. $s$-orbital sites are implemented as single post resonators, while $p_x+i p_y$ sites are realized with three posts in the same resonator, coupled to a Yttrium-Iron-Garnet (YIG) ferrimagnet that energetically isolates the $p_x+i p_y$ mode from the (vestigial) $p_x-i p_y$ and $s$ modes (see Fig.~\ref{fig:LatticeSetup}b, SI Fig.~\SFigref{fig:ChiralSplitting}). The T-symmetry of the ferrimagnet is broken with the $\sim\!0.2$ Tesla magnetic field of $\varnothing 1.6$ mm N52 rare-earth magnets placed outside of the cavity, as close to the YIG as possible to minimize quenching of superconductivity (see SI Fig.~\SFigref{fig:YIGcavityQsA}).

A single fixed-frequency transmon qubit on a sapphire wafer is inserted between the top-left resonator in the $5\times5$ square lattice and the adjacent readout resonator (see Fig.~\ref{fig:LatticeSetup}a); a zoom-in of this setup is shown in Fig.~\ref{fig:LatticeSetup}c.

\section{Probing the Topological Lattice}
\label{Sec:ProbeLattice}

	\begin{figure*}[htp]
		\centering
		\includegraphics[width=\textwidth]{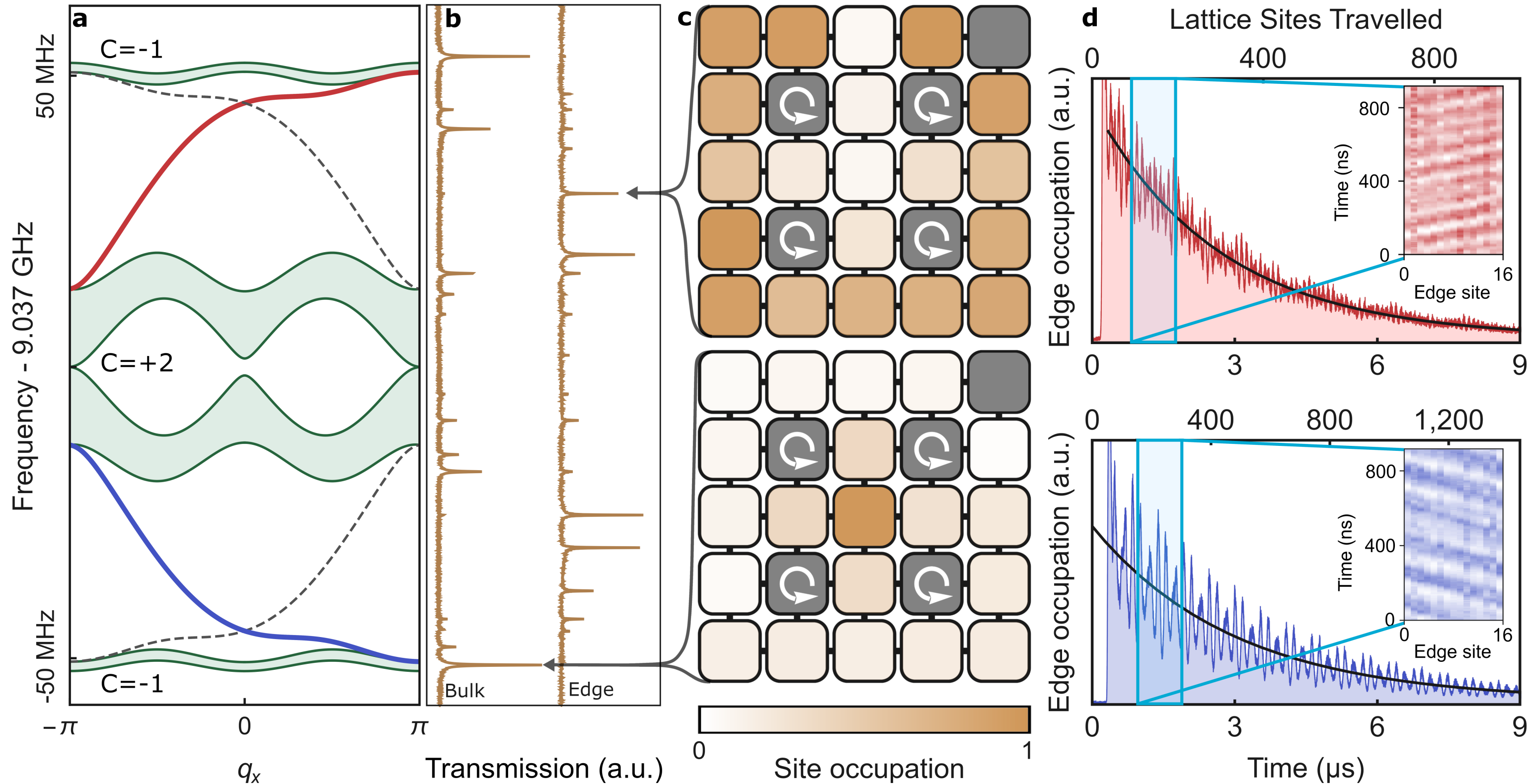} 
		\internallinenumbersJ
		\caption{\textbf{A superconducting Chern circuit.} The central ingredient of a chiral cavity QED platform is a long-lived, spectrally-isolated chiral (unidirectional) mode to couple to a real or synthetic atom. For our experiments this mode is a quantized edge-excitation of a synthetic Hall system realized in a $\alpha=1/4$ Hofstadter square lattice~\protect{\cite{hofstadter1976energy}}. The numerically-computed band structure of our implementation of this model is depicted in \textbf{a,} for an infinite strip geometry. The top and bottom bands each exhibit a Chern number $C=-1$, while the middle two bands, which touch at Dirac points, have a total Chern number $C=+2$; chiral edge channels exist above the bottom band and below the top band, as anticipated from the bulk-boundary correspondence~\protect{\cite{Hasan2010}}. \textbf{b,} shows microwave transmission spectra measured through our actual $5\times 5$ lattice, where both the bulk bands and chiral edges manifest as well-resolved resonant modes due to the finite system size. ``Bulk'' measurements are performed by exciting and measuring at two distinct sites on the interior of the lattice, while ``edge'' measurements employ two sites on the lattice perimeter. In \textbf{c,} we measure the spatial structure of the modes identified with arrows in \textbf{b} and observe that, as expected, the mode residing predominantly in the interior of the lattice is located energetically within a lattice band, while the one localized to the edge resides within an energy gap (see SI~\protect{\ref{SI:LatticeDynamics}} for measurement details). In \textbf{d,} we excite a single edge site at energies within the upper (red) and lower (blue) bulk gaps, and observe the response of the resulting traveling excitation as a function of time averaged over the full perimeter (\textbf{main panels}) and vs. site index around the system edge (\textbf{inset panels}). The insets demonstrate that upper and lower edge channels have opposite chiralities and reflect the numerous orbits of the pulse before it damps away. The ability of a photon to undergo numerous round trips prior to decay is equivalent to spectroscopic resolution of the individual edge modes. \label{fig:SpectroscopyofLattice}}
	\end{figure*}

We first characterize the mode structure of the topological lattice itself in the linear regime, prior to introducing the transmon nonlinearity. Fig.~\ref{fig:SpectroscopyofLattice}a shows the anticipated energy spectrum of a semi-infinite strip $\alpha=1/4$ Hofstadter lattice with four bands and topologically protected edge channels living below the top band and above the bottom band. In a finite system, these continuum bands and edge channels fragment into individual modes satisfying the boundary conditions. Fig.~\ref{fig:SpectroscopyofLattice}b shows the measured response of the lattice when probed \emph{spatially} both within the bulk (left) and on the edge (right), with the energies aligned to Fig.~\ref{fig:SpectroscopyofLattice}a. It is clear that the bulk spectrum exhibits modes within the bands, while the edge spectrum exhibits modes within the bandgaps. We further validate that the modes we have identified as ``bulk'' and ``edge'' modes reside in the correct spatial location by exciting modes identified with arrows in Fig.~\ref{fig:SpectroscopyofLattice}b and performing full microscopy of their spatial structure in Fig.~\ref{fig:SpectroscopyofLattice}c.

To demonstrate that the edge channels are indeed long-lived and chiral (handed), we abruptly excite the system at an edge site within each of the two bulk energy gaps in Fig.~\ref{fig:SpectroscopyofLattice}d (see SI~\ref{SI:LatticeDynamics}). By monitoring the edge-averaged response as the excitation repeatedly orbits the lattice perimeter, we determine that the excitation can circle the full lattice perimeter $>20$ times prior to decay (see Fig.~\ref{fig:SpectroscopyofLattice}d). In the insets to Fig.~\ref{fig:SpectroscopyofLattice}d, we probe in both space \emph{and} time, and observe that the excitations move in opposite directions in the upper and lower band gaps, as anticipated from the bulk-boundary correspondence~\cite{Hasan2010}.

\section{Coupling a Quantum Emitter to the Topological Lattice}
\label{Sec:Qubit}

To explore quantum nonlinear dynamics in the topological lattice we couple it to a transmon qubit (see Fig.~\ref{fig:LatticeSetup} and SI~\ref{SI:TransmonCharacterization}) which acts as a quantized nonlinear emitter whose properties change with each photon that it absorbs. Unlike traditional cavity and circuit QED experiments in which a nonlinear emitter couples to a single mode of an isolated resonator, here the transmon couples to all modes of the topological lattice. In what follows we will induce a controlled resonant interaction between the transmon and individual lattice modes, investigating the resulting strong-coupling physics.

\begin{figure*}[hb]
	 \centering
	 \includegraphics[width=\textwidth]{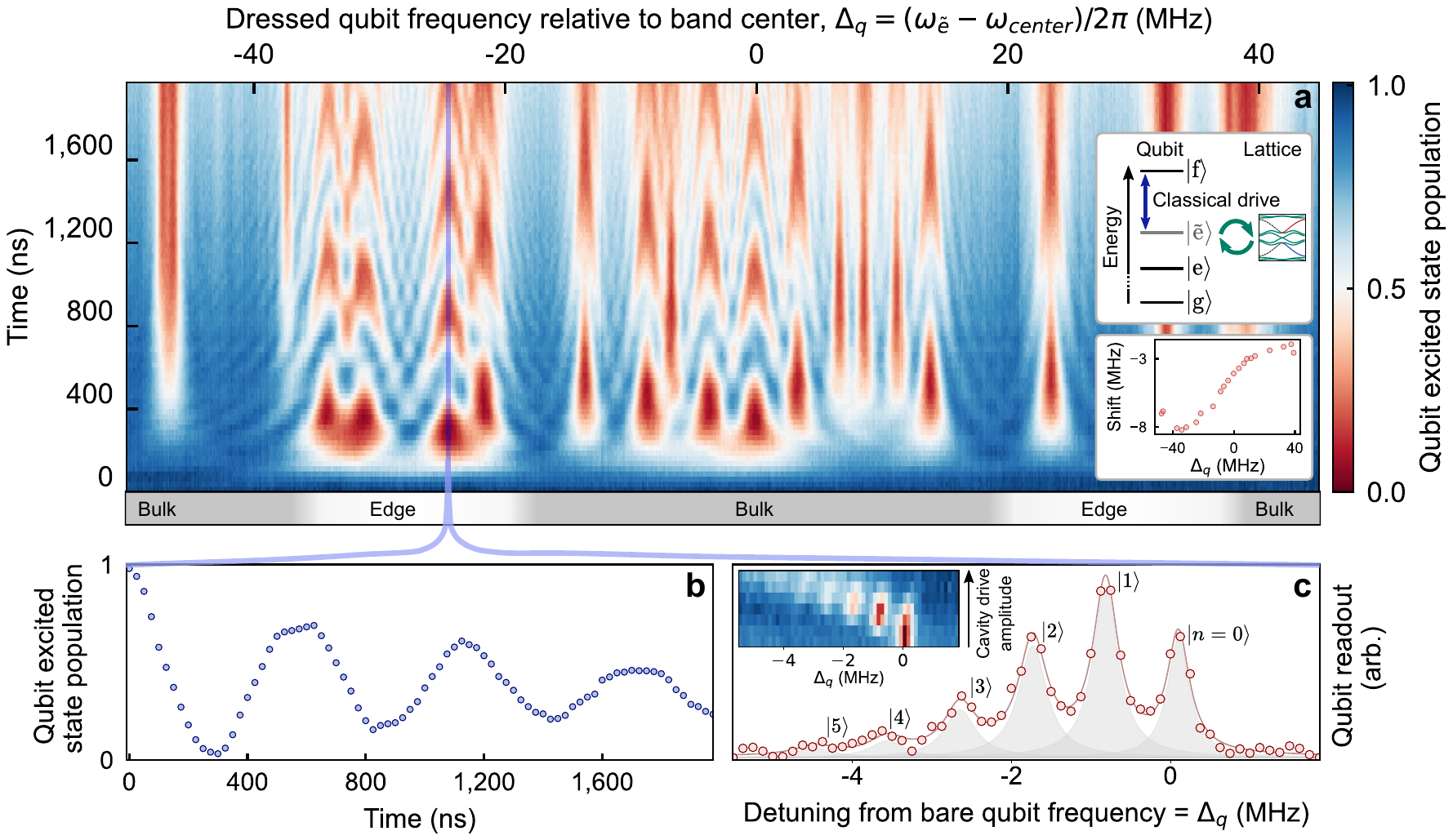}
	 \internallinenumbersJ
	 \caption{\textbf{Quantum nonlinear dynamics on a chiral lattice.} When a transmon qubit is coupled to the edge of the topological lattice, many of the properties of the (nonlinear) qubit are transferred to the (linear) modes of the lattice. In \textbf{a,} we prepare the qubit in its second excited state $|f\rangle$ (see \textbf{top inset}), and drive it with a classical tone (see Methods), thereby scanning the energy of resulting dressed excited state $|\tilde{e}\rangle$ through the lattice band structure. The qubit can then coherently exchange a single photon with the individual lattice modes. The resulting multimode chevron pattern exhibits \emph{fast}, low-amplitude Rabi oscillations when the qubit is detuned from the lattice modes, and slower, high-contrast Rabi oscillations on resonance with each lattice mode. The vacuum Rabi coupling to each mode is determined by the wavefunction overlap of that mode with the qubit site (see SI~\protect{\ref{SI:ModeDependenceShift}}); for this reason all edge modes exhibit fast Rabi oscillations, but many of the bulk modes exhibit slower oscillations. The \textbf{bottom inset} shows the Lamb shift of the transmon due to the topological vacuum of the lattice, measured by comparing the frequency shift of the modes between the chevrons and the linear spectroscopy like Fig.~\protect{\ref{fig:SpectroscopyofLattice}}. There is an additional overall Stark shift of all chevron modes from the strong classical drive. The gray and white at \textbf{bottom} highlight the locations of the bands and gaps. % illustrates the regions of edge and bulk modes in the lattice eigenspectrum through which the qubit is tuned heuristically highlights the . 
	 %In \textbf{b,} we plot the computed band structure through which the qubit is tuned, for comparison; the Lamb shifts are here apparent as a slight compression of the chevron spectrum compared with the computed band structure. 
	 \textbf{b,} When the qubit $|\tilde{e}\rangle$ state is tuned to resonance with a particular lattice mode (blue line in \textbf{a}), vacuum-stimulated Rabi oscillations between qubit and cavity are apparent, demonstrating strong coupling cavity QED where information exchange to a single chiral mode is faster than all decay processes. \textbf{c,} To count photons in a particular edge mode, in this case the highlighted mode in \textbf{a}, we directly excite that mode with a coherent pulse and detect the number of photons it contains as a discrete shift of the qubit frequency (as probed through its readout cavity, see SI~\protect{\ref{SI:TransmonCharacterization}}) resulting from the photon-number-dependent dispersive shift of the qubit frequency. The line is a multi-Lorentzian fit, and the individual Lorentzians for each photon number are shown in gray. \textbf{Inset,} by measuring the qubit excitation spectrum as a function of chiral-mode excitation power, we observe a response transitioning from vacuum (single high-frequency resonance) to the expected Poisson distribution (appearance of multiple lower-frequency resonances). The uncertainties are smaller than the data points in \textbf{b} and \textbf{c} (see SI~\protect{\ref{SI:estimatingerrors}}).
	 }
	 \label{fig:PhotonCounting}
	\end{figure*}

The $|g\rangle \leftrightarrow |e\rangle$ transition of the transmon ($\omega_{q}\approx 2\pi\times 7.8$ GHz) is detuned from the lattice spectrum ($\omega_\text{lattice}\approx 2\pi\times 9$ GHz) by $\Delta\approx 2\pi\times 1.2$ GHz. We bring the transmon controllably into resonance with individual lattice modes via the dressing scheme in Fig.~\ref{fig:PhotonCounting}a inset (see Methods and ref.~\cite{pechal2014microwave} for details); this dressing also gives us complete control over the magnitude of the qubit-lattice site coupling.

In Fig.~\ref{fig:PhotonCounting}a we tune the excited transmon into resonance with individual lattice modes and observe vacuum-stimulated Rabi oscillations (see SI ~\ref{SI:PulsedStructure}) of a quantized excitation between the transmon and the mode. Comparing with the predicted band structure shown in Fig.~\ref{fig:PhotonCounting}b, we see that the transmon couples efficiently to both bulk and edge modes of the lattice, despite being physically located on the edge. This is because the lattice is only 5 sites across, comparable to the magnetic length $l_B\sim 1/\alpha=4$ sites, so the lattice site coupled to the transmon has substantial participation in both bulk and edge modes; furthermore, the system is sufficiently small that the number of bulk sites is comparable to the number of edge sites, so all modes have approximately the same ``volume.''

To unequivocally demonstrate strong coupling between the transmon and a single lattice mode, we examine a single frequency slice of Fig.~\ref{fig:PhotonCounting}a versus evolution time. Fig.~\ref{fig:PhotonCounting}c shows such a slice and demonstrates high-contrast oscillations that take several Rabi cycles to damp out, as is required for strong light-matter coupling. For simplicity, we choose our dressed coupling strength to be less than the lattice mode spacing; stronger dressing to explore simultaneous coupling to multiple lattice modes opens the realm of super-strong-coupling physics~\cite{meiser2006superstrong,kuzmin2019superstrong}, where the qubit launches wavepackets localized to smaller than the system size.

When a qubit is tuned towards resonance with a single cavity mode it experiences level repulsion~\cite{fragner2008resolving} and then an avoided crossing at degeneracy. The situation is more complex for a qubit coupled to a full lattice,  where one must account for interactions with \emph{all} lattice modes, both resonant and non-resonant. In total these couplings produce the resonant oscillations observed in Fig.~\ref{fig:PhotonCounting}c plus a frequency-dependent shift due to level repulsion from off-resonant lattice modes, which may be understood as a Lamb shift from coupling to the structured vacuum~\cite{lamb1947fine}. We quantify this Lamb shift by comparing the frequencies of the modes observed in linear lattice spectroscopy, as in Fig.~\ref{fig:SpectroscopyofLattice}a but with the transmon present (see SI~\ref{SI:LatticeSpec}), to those observed in chevron spectroscopy in Fig.~\ref{fig:PhotonCounting}a. These data are shown in the lower inset to Fig.~\ref{fig:PhotonCounting}a. When the qubit is tuned near the low-frequency edge of the lattice spectrum it experiences a downward shift from all of the modes above it, and when it is tuned near the upper edge of the lattice, it experiences a corresponding upward shift. These two extremes smoothly interpolate into one another as modes move from one side of the qubit to the other. There is also a near-constant Stark shift of $~\sim$ 3.5 MHz arising from the classical dressing tone. To our knowledge, this is the first measurement of the Lamb shift of a qubit in a synthetic lattice vacuum.

Finally, we demonstrate the ability to count photons within an individual lattice mode. If the transmon were coupled to a single lattice site and not to the full lattice, each photon in that site would shift the qubit $|g\rangle\leftrightarrow|e\rangle$  transition by $2\chi$, where $\chi\approx \frac{g_l^2}{\Delta}\times\frac{\alpha_q}{\Delta+\alpha_q}\approx 2\pi\times 5$ MHz,  and $\alpha_q$ is the transmon anharmonicity. This photon-number-dependent shift, and thus the intra-cavity photon number, can be measured by performing qubit spectroscopy detected through the readout cavity (see SI ~\ref{SI:ModeDependenceShift}). When the transmon is coupled to a lattice rather than an isolated cavity, the $\chi$ shift is diluted by the increased volume of the modes. In Fig.~\ref{fig:PhotonCounting}d, we inject a coherent state into the highlighted mode in Fig.~\ref{fig:PhotonCounting}a and then perform qubit spectroscopy to count the number of photons within the mode. The observed spectrum corresponds to a coherent state with $\bar{n}\approx 1.4$, with the individual photon occupancies clearly resolved. Indeed, when we perform this experiment as a function of the amplitude of the coherent excitation pulse (Fig.~\ref{fig:PhotonCounting}d, inset), we find a continuous evolution from vacuum into a Poisson distribution over the first six Fock states.

%The relative couplings can be extracted from measured dispersive shifts shown in Fig.~\ref{fig:PhotonCounting}c
	
\section{Outlook}
\label{Sec:Conclusion}
	
In this work we have demonstrated a photonic materials platform that combines synthetic magnetic fields for lattice-trapped photons with a single emitter. This has enabled us to explore interactions between the individual modes of a topological system and the non-linear excitation spectrum of the emitter, entering for the first time the realm of fully-granular chiral cavity QED and thus demonstrating the ability to count and manipulate individual photons in each mode of the lattice. We anticipate that coupling a transmon to a longer edge would enable qubit-mediated photon-induced deformation of the edge channel (in the ``super strong'' coupling limit of the edge channel~\cite{meiser2006superstrong,sundaresan2015beyond}), as well as universal quantum computation via time-bin-encoding~\cite{Pichler11362} or blockade engineering~\cite{chakram2020multimode}. Introduction of a qubit to the bulk of this system would allow investigation of the shell-structure of a Landau-photon polariton~\cite{de2020light}, a precursor to Laughlin states. Addition of a second qubit on the edge would allow chiral, back-scattering-immune quantum communication between the qubits~\cite{lodahl2017chiral}. Scaling up to one qubit per site will enable dissipative stabilization~\cite{ma2017autonomous,kapit2014induced, PhysRevB.92.174305,lebreuilly2017stabilizing} of fractional Chern states of light~\cite{anderson2016engineering} and thereby provide a clean platform for creating anyons and probing their statistics~\cite{wilczek1990fractional}.

\section{Methods}
%topics to cover:
%%how does the dressing that makes the transmon resonant with different modes work
%%how large is our lattice-mode anharmonicity from first principles

The transmon qubit (see SI~\ref{SI:Transmon Fab}) has a $|g\rangle \leftrightarrow |e\rangle$ transition frequency of $\omega_q=2\pi\times 7.8$ GHz, compared with the lattice spectrum centered on $8.9$ GHz and spanning $\pm 50$MHz (due to lattice tunneling $J=2\pi\times 18$MHz). In the presence of the applied magnetic fields, the $|g\rangle \leftrightarrow |e\rangle$ transition of the transmon exhibits a $T_1=2.9\mu$s and $T_2=3.9\mu$s (see SI~\ref{SI:TransmonCharacterization}). The anharmonicity of the transmon is $\alpha_q=2\pi\times 346$ MHz. The transmon is coupled to its readout cavity ($\omega_r=2\pi\times 10.6$ GHz, $\kappa_r=2\pi\times 500$ kHz) with $g_r=2\pi \times 66$ MHz. The transmon is coupled to the $(1,1)$ site of the lattice with $g_l=2\pi\times 111$MHz. The $s$-like lattice sites have a linewidth of $\kappa_s=2\pi\times 5$kHz, while the $p_x+i p_y$-like sites have a linewidth $\kappa_{p_x+i p_y}=2\pi\times 50$ kHz. The lattice sites themselves are tuned with $\pm 1$ MHz accuracy for the $s$-sites and $\pm 3$ MHz accuracy for the $p_x+i p_y$ sites.

We perform linear spectroscopy of the lattice with dipole antennas inserted into each lattice site and use cryogenic switches to choose which sites we excite/probe (see SI~\ref{SI:Lattice Fab}).

We perform nonlinear spectroscopy by exciting the transmon through its readout cavity. The transmon is fixed-frequency to avoid unnecessary dephasing from sensitivity to the magnetic fields applied to the YIG spheres. As a consequence, we ``tune'' the transmon to various lattice modes by dressing through the readout cavity~\cite{pechal2014microwave}: we prepare the transmon in the second excited ($|f\rangle$) state and then provide a detuned drive on the $|f\rangle\leftrightarrow |e\rangle$ transition to create a dressed $|\tilde{e}\rangle\approx|f\rangle-\frac{\Omega}{\Delta}|e\rangle$ state at any energy in the vicinity of the lattice band structure, with a dipole moment for coupling to the lattice which is rescaled by the ratio of the dressing Rabi frequency to the detuning from the $|f\rangle\leftrightarrow |e\rangle$ transition. The resulting vacuum-stimulated $|f,0\rangle\leftrightarrow |g,1_k\rangle$ Rabi frequency $g_k\approx g_l\times\frac{\Omega}{\Delta}\times\langle x_\text{transmon}|\psi_k\rangle$. Here $\Omega$ is the Rabi frequency of the dressing tone on the $|f\rangle\leftrightarrow |e\rangle$ transition of the transmon and $\langle x_\text{transmon}|\psi_k\rangle$ is the participation within mode $k$ of the lattice site where the transmon resides. The dressing scheme may alternatively be understood as a 2-photon Rabi process, where the $|f,0\rangle\leftarrow |e,0\rangle$ transition is stimulated by the classical drive, and the $|e,0\rangle\leftarrow |g,1_k\rangle$ transition is stimulated by the vacuum field of mode k.

For the qubit measurements, the the lattice is tuned to a center frequency of $2\pi\times 8.9$ GHz, corresponding to a dressing frequency of $2\pi\times 6.35$ GHz $\pm 50$ MHz. Note that with the additional significant figure, the $|g\rangle\leftrightarrow|e\rangle$ transition has a frequency of $2\pi\times 7.75$ GHz.

\section{Acknowledgements}
This work was supported primarily by ARO MURI W911NF-15-1-0397 and AFOSR MURI FA9550-19-1-0399. This work was also supported by the University of Chicago Materials Research Science and Engineering Center, which is funded by National Science Foundation under award number DMR-1420709. J.O., M.P., and G.R. acknowledge support from the NSF GRFP. We acknowledge Andrew Oriani for providing a rapidly cycling refrigerator for cryogenic lattice calibration.

\section{Author Contributions}
The experiments were designed by R.M. J.O., D.S. and J.S. The apparatus was built by J.O., R.M., and B.S. J.O. and M.P. collected the data, and all authors analyzed the data and contributed to the manuscript.
	
\section{Author Information}
The authors declare no competing financial interests. Correspondence and requests for materials should be addressed to D.S. (dis@uchicago.edu).
	
\section{Data Availability}
The experimental data presented in this manuscript is available from the corresponding author upon request.

\bibliographystyle{naturemag}
\bibliography{ChiralCavityQED}

\renewcommand{\tocname}{Supplementary Information}
\renewcommand{\appendixname}{Supplement}
	
\setcounter{equation}{0}
\renewcommand{\theequation}{S\arabic{equation}}
\renewcommand{\thefigure}{S\arabic{figure}}

\incltocpage
\clearpage

%\tableofcontents
\appendix
\setcounter{secnumdepth}{2}

\section{Cryogenic Setup}
\label{SI:DilFridgeWiring}
% SI sections- fridge wiring/physical setup

% Separate figure counter for the Supplement
\newcounter{sfigure}
\setcounter{sfigure}{1}
\renewcommand{\thefigure}{S\arabic{sfigure}}

\begin{figure}
 \includegraphics[width=0.48\textwidth]{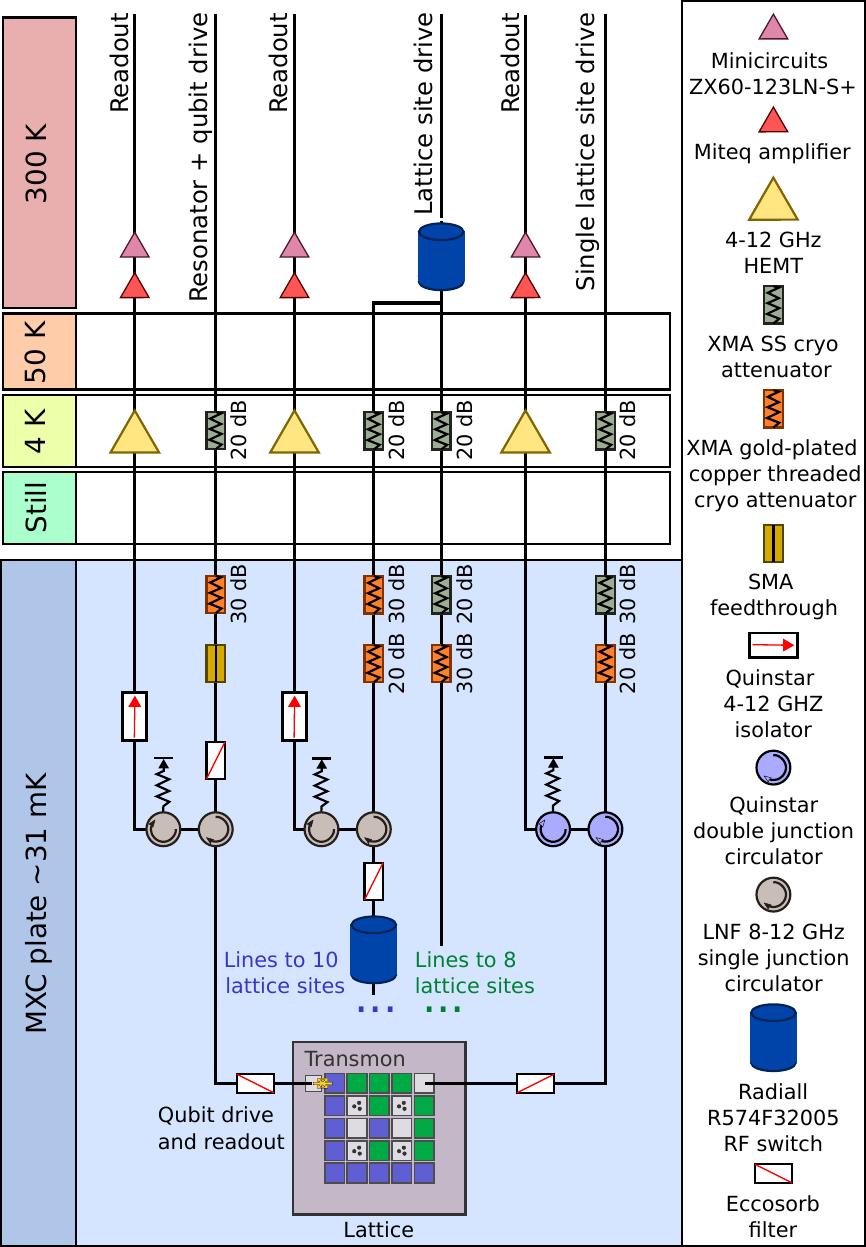}
 \internallinenumbersJ
 \caption{ \textbf{Partial diagram of cryogenic setup.} Experiments involving the qubit were performed in a Bluefors LD-250 dilution refrigerator operating at approximately 31 mK. The antennas coupled to some lattice sites (highlighted in blue) are accessible via an input and readout line connected by a circulator, allowing reflection measurements on individual sites. Other antennas coupled to lattice sites (highlighted in green) are accessible via input line only. 
 \label{fig:SetupFigWiring}
 }
\end{figure}

The cryogenic setup employed for measuring the lattice coupled to the qubit is shown in Fig.~\SFigref{fig:SetupFigWiring} and is similar to the setup used to measure the lattice prior to the introduction of the qubit. The lattice is mounted on the mixing chamber (MXC) plate of a Bluefors LD-250 dilution refrigerator at $\sim\!31$ mK. The lattice sites depicted in blue are connected to a 10-way cryogenic switch which is in turn connected to a circulator so that these sites can all connect to either an input line or an output line. Sites in green are connected to only input lines and thus can only be excited, not measured. The site $(1,1)$ is connected separately to a circulator so that it can be measured and excited independently of the blue lattice sites. The remaining six sites in the 5x5 lattice are not connected to a either an input line or output line. The qubit readout resonator is also connected to a circulator so that reflection measurements can be performed to measure the state of the qubit. The qubit is excited off-resonantly through the readout resonator. Each output line is filtered by an eccosorb filter to suppress accidental high-frequency qubit excitation.

For the measurements performed without a qubit (see Fig.~\ref{fig:SpectroscopyofLattice}), \emph{two} cryogenic switches were used within the fridge, enabling direct probing of 20 sites, while a 21st site was independently connected to a circulator. The only sites not measured were the 4 chiral cavities, whose modes are more localized at the bottom of the cavity, making coupling to them via a dipole antenna difficult without spoiling the quality factor. 

\begin{figure}
    \centering
    \includegraphics[width=0.48\textwidth]{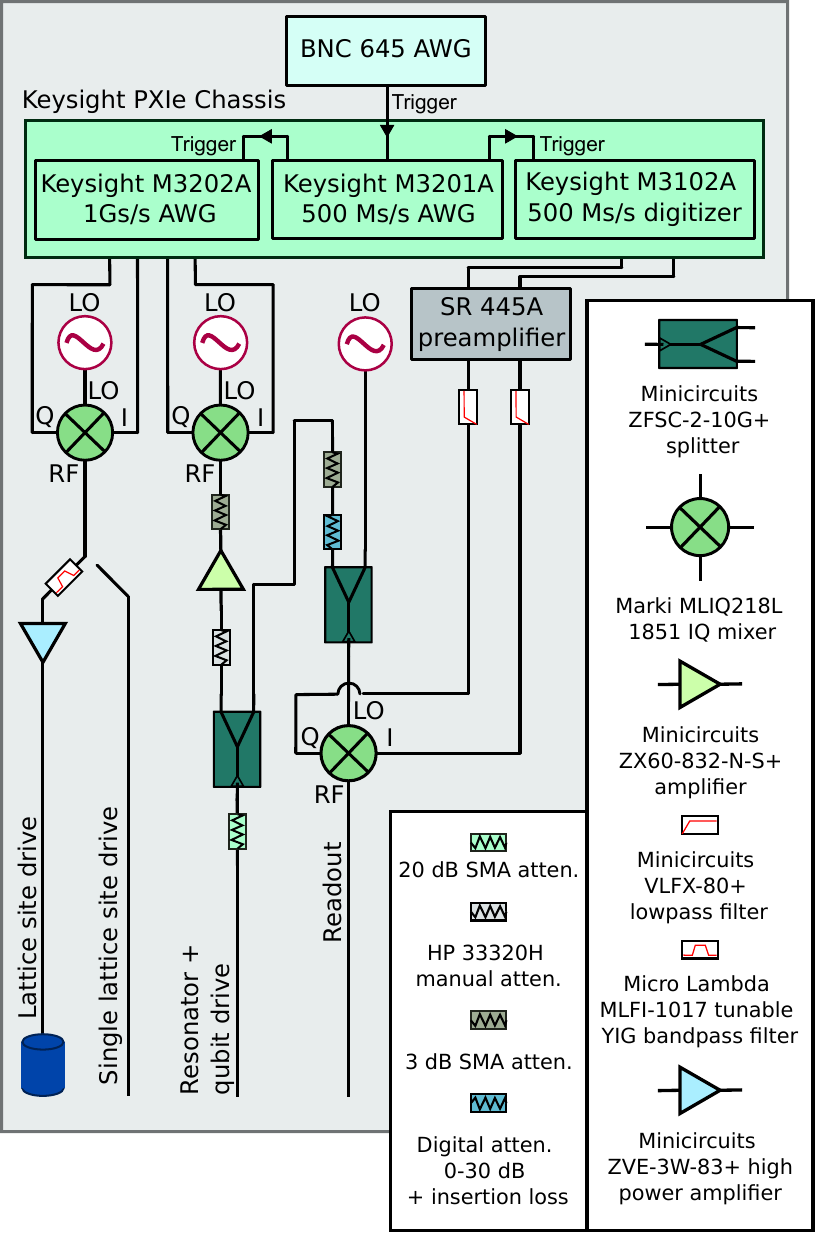}
    \internallinenumbersJ
    \caption{\textbf{Schematic of room-temperature experimental setup.} In any given measurement, one of the three readout lines pictured in Fig.~\SFigref{fig:SetupFigWiring} is selected for use. Drives on lattice sites, important for the pulsed measurements with sequences detailed in Fig.~\protect{\SFigref{fig:PulseChains}}, are either passed directly into the refrigerator after I/Q modulation or, in the case of strong dressing tones, are filtered by a YIG (Yttrium Iron Garnet) tunable bandpass filter and amplified before use.}
    \label{fig:SetupFigRoomTemp}
\end{figure}

A Keysight PNA-X N5242 is used to perform lattice spectroscopy. For the qubit measurements, a Keysight arbitrary waveform generator (M8195A, $64$ GSa/s) is used to synthesize a local oscillator signal near the qubit frequency, while Berkeley Nucleonics 845-M microwave synthesizers provide separate local oscillator signals near lattice and dressing frequencies. The local oscillators are then I/Q modulated by Keysight PXIe AWGs (M3202A, multichannel, 1GS/s) to generate the individual qubit drive, qubit readout, and dressing pulses. The qubit drive and readout pulses are combined outside the fridge and sent to the readout resonator (Fig.~\SFigref{fig:SetupFigRoomTemp} and Fig.~\SFigref{fig:SetupFigWiring}). The reflected readout signal is routed to the output line via circulators and amplified with a HEMT amplifier at 4 K and additional room temperature amplifiers (Miteq AFS3-00101200-22-10P-4, Minicircuits ZX60-123LN-S+). The signal is then demodulated using an IQ mixer and recorded using a fast digitizer (Keysight M3102A, 500 MSa/s). A schematic layout of the room-temperature components of the experimental setup can be found in Fig.~\SFigref{fig:SetupFigRoomTemp}.

\setcounter{sfigure}{2}

\section{Transmon Fabrication}
\label{SI:Transmon Fab}

The transmon qubit is made of aluminum deposited on a \o$50.8$mm, 430$\mu$m thick C-place (0001) sapphire wafer. The wafer was annealed at $1200^\circ$ C for 1.5 hours. The wafer was then cleaned with toluene, acetone, isopropanol, and DI water in an ultrasonic bath immediately prior to junction deposition. The transmon was deposited in a single layer. The mask for deposition was defined by electron-beam lithograthy using a Raith EBPG5000 Plus E-Beam Writer. The mask was a bi-layer of resist made of a stack of MMA and PMMA. The capacitor pads were rectangles with dimensions 50 microns by 1100 microns and were written with the 40 nA beam. The junctions were patterned Manhattan-style using a finer $1$ nA beam. In the intermediate region that connects the junction to the capacitor a beam of $4$ nA was used. Before writing, a 50 nm layer of Au was thermally evaporated onto the wafer in order to provide adequate grounding the the electron beam. After writing the resist, the resist stack was developed for 1.5 minutes in a solution of 3 parts IPA and 1 part DI water chilled on a 6$^\circ$ C cold plate. The junction was then deposited using electron-beam evaporation in three steps. First, 80 nm of aluminum was deposited at an angle of $40^\circ$ relative to the plane of the wafer and parallel to a finger of the junction. Next, the junction was oxidized in $20$ mbar of high purity oxygen for 10 minutes. Last, a 45 nm layer of aluminum was deposited at the same evaporation angle of $40^\circ$ but orthogonal to the first evaporation. After evaporation, the rest of resist as well as the aluminum attached to the resist were removed via lift-off in an $80^\circ$ C solution of PG remover for 3 hours. The wafer was thereafter diced into the dimensions that fit the lattice setup. 

\section{Lattice Fabrication}
\label{SI:Lattice Fab}

The lattice is machined from a solid block of niobium, which exhibits a low $H_{c1}$ to allow magnetic fields to penetrate the material, and a high $H_{c2}$ to ensure that the field does not quench the superconductivity~\cite{karasik1970superconducting}. High niobium purity is not required since the magnetic field used to bias the YIG spheres already limits the quality factor of the cavities. The resonators are arranged in a $5\times 5$ square lattice, which is the minimum lattice size that supports a clear distinction between bulk and edge. While the edge channels predominantly reside in sites on the lattice perimeter, they have some participation in the sites one removed from the edge, exponentially decaying into the bulk. Each lattice site consists of a $\frac{\lambda}{4}$ coaxial resonator described in detail in our earlier work~\cite{Owens2018}. The frequency of the lattice site is inversely dependent on the length of the post in the center of the cavity. By adding indium foil to the end of a post the site frequency can be tuned to a precision better than 1 MHz. In order to maximize the quality factor of the resonator, the depth of the cavity is set so that the evanescent decay from the post mode is much less than the residual resistive loss of the superconductor. The niobium lattice is mounted in a copper box that is screwed into the niobium in order to adequately thermalize the niobium to the mixing chamber plate of the dilution refrigerator in which it is placed. Antennas are mounted onto the lid of the copper box so that a single antenna protrudes into each lattice site from the top of the cavity. The length of the antenna sets the coupling quality factor of each lattice site. For these measurements, each lattice site is weakly coupled to its antenna so that the total quality factor of the lattice is maximized. 

Special care is required to maintain cryo- and qubit compatibility with the ferrimagnetic elements and their associated B-fields. The YIG spheres are located physically inside of their lattice cavities in order to couple to the microwave modes, so magnetic field must be routed to the YIG spheres through bulk superconductor. We make the cavities out of the type II superconductor niobium so that magnetic fields can penetrate it while it is in the vortex superconducting state. To create the bias field, we place a 1.6 mm diameter permanent neodymium cylindrical magnet in a hole outside of the cavity but directly underneath each YIG sphere, leaving a 0.3 mm thick layer of niobium between the magnet and the inside of the cavity and the YIG (Fig.~\ref{fig:LatticeSetup}c). The proximity of the magnet to the YIG sphere achieves a bias field of $\sim\!0.2$ T on the YIG sphere, while the small size of the magnet minimizes the amount of field that passes through the posts of the cavity and the amount of normal or vortex state niobium in areas with large current flow. This bias field achieves splittings between the two chiral rotating modes of the YIG cavity of $2\pi\times 200$ MHz while retaining cavity quality factors of $2\times 10^5$. Transmission through the two chiral modes of the YIG cavity is shown in Fig.~\SFigref{fig:ChiralSplitting}.

Cavity quality factors were measured after machining the lattice, inserting the ferrites (YIG spheres) to relevant cavities, and applying magnetic fields. After machining but before applying any surface treatment, lattice sites had quality factors of $\sim 2\times 10^6$. Introducing the YIG sphere to the cavity does not degrade the Q, so long as no additional materials are added to hold the YIG sphere in place. After applying the magnetic field in the final configuration, cavity quality factors dropped to $\sim 2\times 10^5$, suggesting that the limiting loss factor of the cavity modes is the resistive losses in the normal regions created by the magnetic field piercing the superconductor.

The cavities are coupled together via holes milled into the back side of the niobium block. These holes open paths between lattice sites that allow the Wannier functions of the lattice sites to overlap with their neighbors, creating coupling between sites. These coupler holes create additional coupling via a virtual coupling mechanism, in which the couplers act as higher frequency resonators. In some (non-cryogenic) lattices we added a screw that could tune the frequency of the coupler lower, allowing us to achieve greater couplings between lattice sites (up to 150 MHz). For the cryogenic lattice design we reduced the tunneling between sites to $J=2\pi\times 18$ MHz. This was done because we wanted to preserve the quality factor of the lattice modes by using less magnetic field, which in turn reduced the amount by which we broke time-reversal symmetry. In order to increase both the quality factor of the modes and the tunneling ratio, more effective methods of funneling magnetic field are required.

Three readout cavities for qubits are machined on the edge of the lattice, though for the results shared in this letter only a single qubit was introduced to one such cavity. These cavities are designed to be much higher in frequency ($\sim10.5$ GHz) than the main lattice so that they do not interact directly with it. The qubit is mounted so that its capacitor pads act as antennas that can directly couple the qubit both to the readout cavity and to the lattice modes (see Fig.~\ref{fig:LatticeSetup}d). The antenna that couples to the readout cavity is inserted into the bottom of this cavity so that it can attain a low coupling Q (20,000), while being short enough that the modes introduced by the antenna are much higher in frequency than the readout cavity.

\section{Cryogenic Chiral Lattice Site Characterization}
\label{SI:CryoLatticeData}
\begin{figure}
	 \centering
	 \includegraphics[width=.48\textwidth]{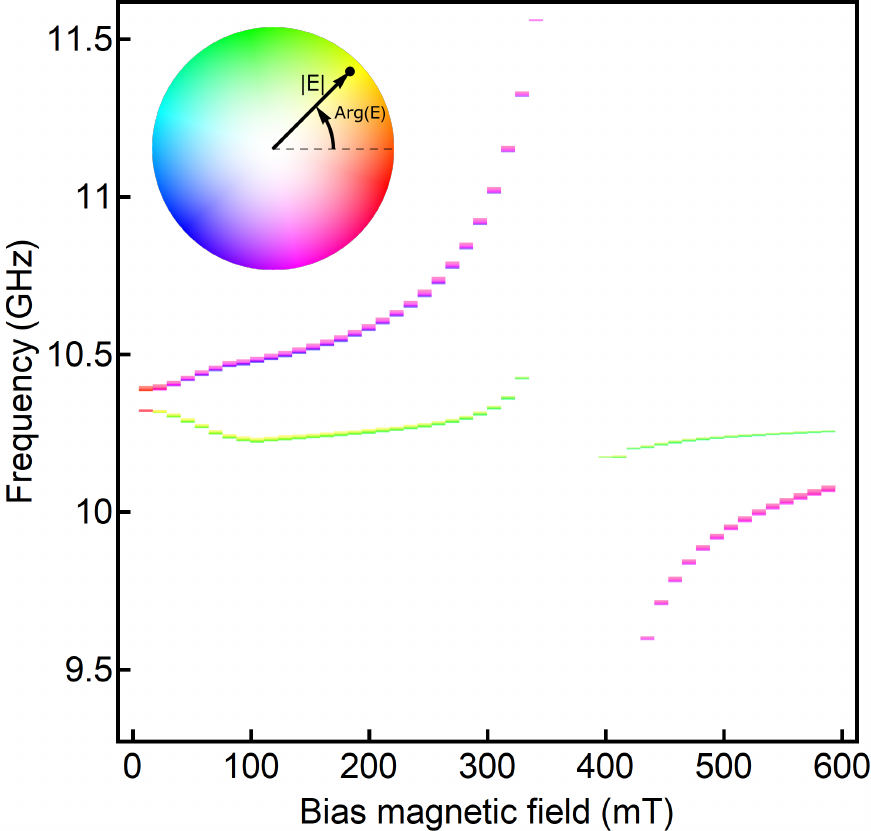}
	 \internallinenumbersJ
	 \caption{\textbf{Splitting $p_x\pm i p_y$ modes in a B-field.} Chiral cavity mode frequencies are measured as a function of magnetic field applied to the YIG sphere. For each slice in the y-direction, the cavity is warmed above the $T_C$ of niobium to 12 K, so that the magnetic field on the YIG sphere can be changed. The cavity is then cooled to 2 K and the magnetic flux is locked in place by the superconducting transition of the cavity. Transmission is then measured between two antennas placed $45^\circ$ apart (with respect to the center of the cavity) so that the phase accumulated in transmission can be measured. In this plot, we take the difference between transmission from antenna 1 to 2 and transmission from antenna 2 to 1 to isolate the non-reciprocal phase shift in the cavity. This plot shows both the magnitude and phase of transmission; the color shows the phase while the brightness shows the magnitude. The two cavity modes split in frequency when the magnetic field is applied and the phase shifts of the modes are $-90^\circ$ and $+90^\circ$ due to the opposite chirality of the modes.}
	 \label{fig:ChiralSplitting}
	\end{figure}

A major innovation of this work was its design of a way to break time-reversal symmetry in the lattice while maintaining low-loss modes and compatibility with qubits that are sensitive to magnetic fields. In prior work we designed interactions between 3D microwave photons and a DC magnetic field in a room temperature aluminum lattice~\cite{Owens2018,anderson2016engineering}. This kind of interaction is mediated via a ferrite sphere (made of Yttrium Iron Garnet, or YIG) placed inside a lattice cavity. A DC magnetic field is applied to the YIG sphere through the cavity, tuning magnon modes of the sphere into resonance with the cavity modes. The hybridization of the chiral magnon modes with the 3D cavity modes results in time-reversal-symmetry-broken cavity modes. Lattice cavities populated with YIG spheres are engineered to host two degenerate modes. One of the modes' magnetic fields at the center of the cavity precesses clockwise, and the other precesses counter-clockwise. This opposite chirality of the modes creates a difference in the coupling strength between the two cavity modes and the chiral YIG mode. The mode that precesses with the same chirality as the YIG sphere couples more strongly to the chiral YIG mode than the mode with the opposite chirality. In Fig.~\SFigref{fig:ChiralSplitting}, we show the mode frequencies of the cavity as a function of an applied uniform magnetic field on a cavity with a YIG sphere inside. At low fields, the two cavity modes are nearly degenerate, but as the bias magnetic field is increased, the YIG mode comes into resonance with the cavity and splits the cavity modes by up to 400 MHz. The color of plotted data indicates the phase acquired during transmission through the cavity via ports that are placed $45^{\circ}$ apart with respect to the cavity center. Transmission through one of the chiral modes results in a phase accumulation of $\frac{\pi}{2}$ radians, while the other chiral mode acquires a phase of $\frac{3\pi}{2}=-\frac{\pi}{2}$. The observed phase is twice the physical angle between the two ports because we are comparing $S_{21}$ and $S_{12}$ to remove global reciprocal phases accrued in cables. 

\setcounter{sfigure}{3}

\begin{figure}
 \includegraphics[width=.48\textwidth]{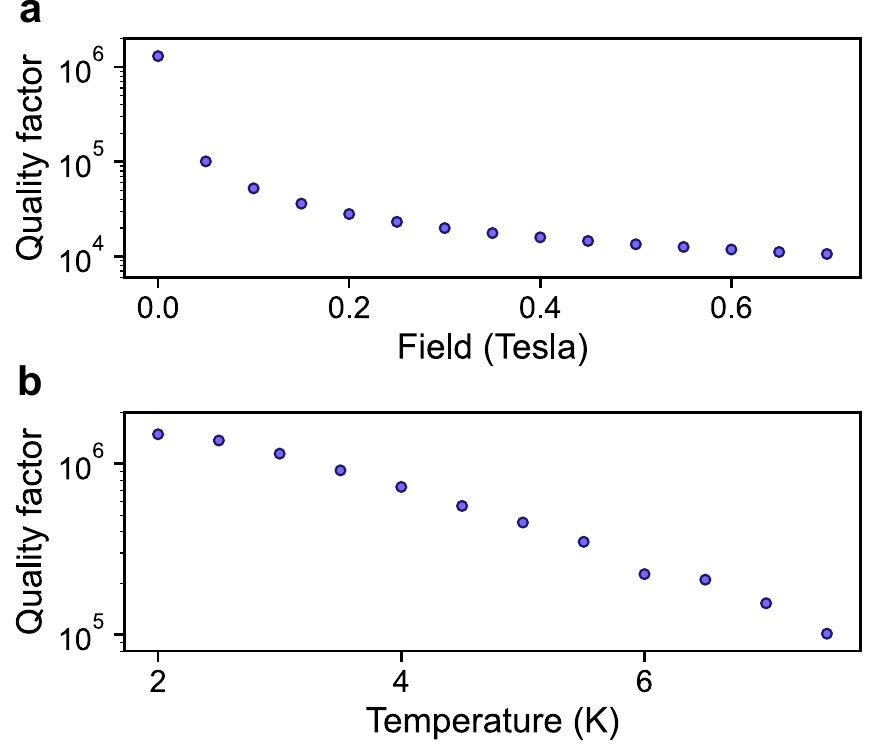}
 \internallinenumbersJ
 \caption{\textbf{Quality factor of chiral cavities.} \textbf{a,} Quality factor of a three post cavity as a function of magnetic field. For every measurement, the cavity is warmed above niobium's superconducting transition of 9 K to 12 K so that the magnetic field can be changed. The cavity then is cooled to 2 K within the magnetic field and the quality factor is measured. The cavity quality factor decreases to 30000 at the field that tunes the YIG to the cavity resonance. \textbf{b,} Quality factor of a three post cavity with a YIG sphere as a function of temperature while the magnetic field is held at zero. The YIG sphere lowers the quality factor of the cavity but the quality factors still reach ${\raise.17ex\hbox{$\scriptstyle\sim$}} 2$ million.
 \label{fig:YIGcavityQsA}
 }
\end{figure}

In practice, routing the magnetic field to the ferrite poses a challenge when using superconducting lattices, as magnetic fields either are repelled by superconductors or induce normal regions that greatly increase the cavity loss. In Fig.~\SFigref{fig:YIGcavityQsA} we show the effects of an increasing magnetic field applied to a niobium cavity. The field is uniform, applied externally at $12$ Kelvin (K) before the cavity superconducts. For each data point the cavity is warmed to 12 K before the field is changed. At the fields used to bias the YIG sphere to the cavity frequency (350 mT as shown in Fig.~\SFigref{fig:ChiralSplitting}), the quality factor of the cavity would be as low as $10^4$. Additionally, the Josephson junction in the qubit is made out of aluminum, a type I superconductor with a low $H_{c1}$~\cite{cochran1958superconducting}. Simply applying an external magnetic field would adversely effect both qubit and cavity lifetimes. 

\setcounter{sfigure}{4}

\begin{figure}
 \includegraphics[width=.48\textwidth]{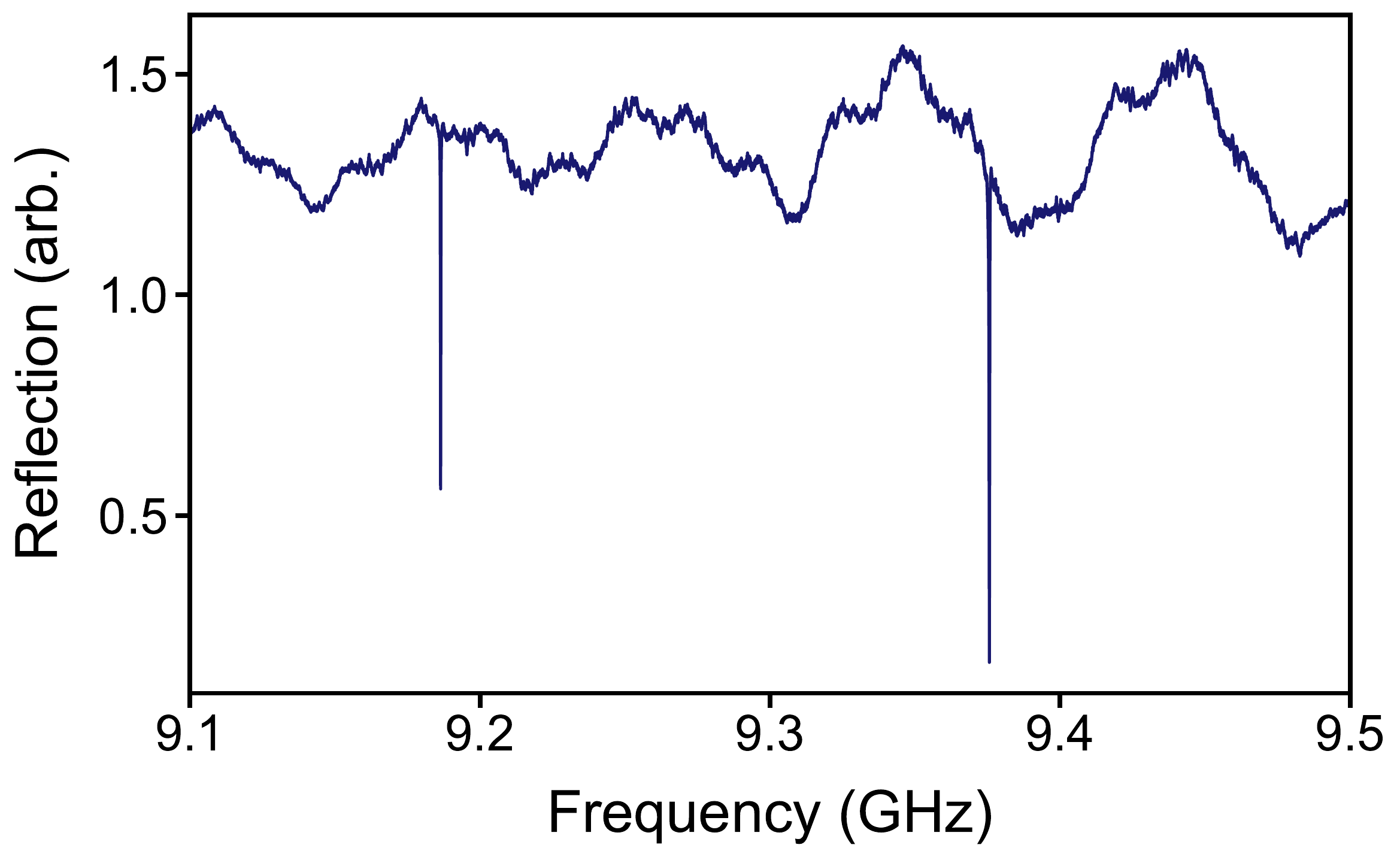}
 \internallinenumbersJ
 \caption{\textbf{Chiral cavity spectrum with a permanent magnet.} The magnet is inserted into a hole bored from the bottom of the cavity so that the magnet sits below the YIG sphere with $\sim$ 0.5 mm of niobium between the magnet and the YIG sphere. The ensemble is then cooled to 2 K. The magnet generates enough B-field to break the degeneracy of the two chiral modes ($p_x\pm i p_y$) by $\sim\!200$ MHz while still maintaining a quality factor of $2\times 10^5$.}
 \label{fig:YIGcavityQsB}
\end{figure}

To minimize the magnetic fields permeating the system, we generate a \emph{local} magnetic field in the vicinity of the YIG spheres with small Neodymium magnets ($\sim1.5$mm diameter). In order to achieve the required field at each ferrite, we insert a magnet into a small hole milled into the back the cavity (shown in Fig.~\ref{fig:LatticeSetup}) directly beneath the ferrite. This minimizes the amount of niobium which sees its superconductivity quenched by the B-field, as the strongest part of the magnetic field is localized to the area between the ferrite and the magnet: indeed, the bulk of the modal surface currents flows in the posts of the cavity, locations where the B-field has decayed substantially. As shown in Fig.~\SFigref{fig:YIGcavityQsB}, we generate sufficient field to break the degeneracy between the two chiral cavity modes by $200$ MHz, whilst maintaining a quality factor of 200,000. The splitting between these two modes is a measure of how strongly time-reversal symmetry is broken. This splitting also limits the maximum tunneling rate in the lattice, as the Hofstadter model assumes one orbital per lattice site, requiring the tunneling energy to remain small enough to avoid coupling to the counter-rotating orbitals. The ratio between the tunneling rate and the loss rate in our cavities is then a measure of how fast the dynamics are compared to the loss rate, which is an important benchmark for the system that determines how far the photons move within their lifetimes. In this work, the ratio between tunneling and loss rates is ~$\frac{18 \textrm{MHz}}{50\textrm{kHz}}\approx 400$.

\setcounter{sfigure}{5}

\setcounter{sfigure}{6}
\begin{figure}
    \centering
    \includegraphics[width=.47\textwidth]{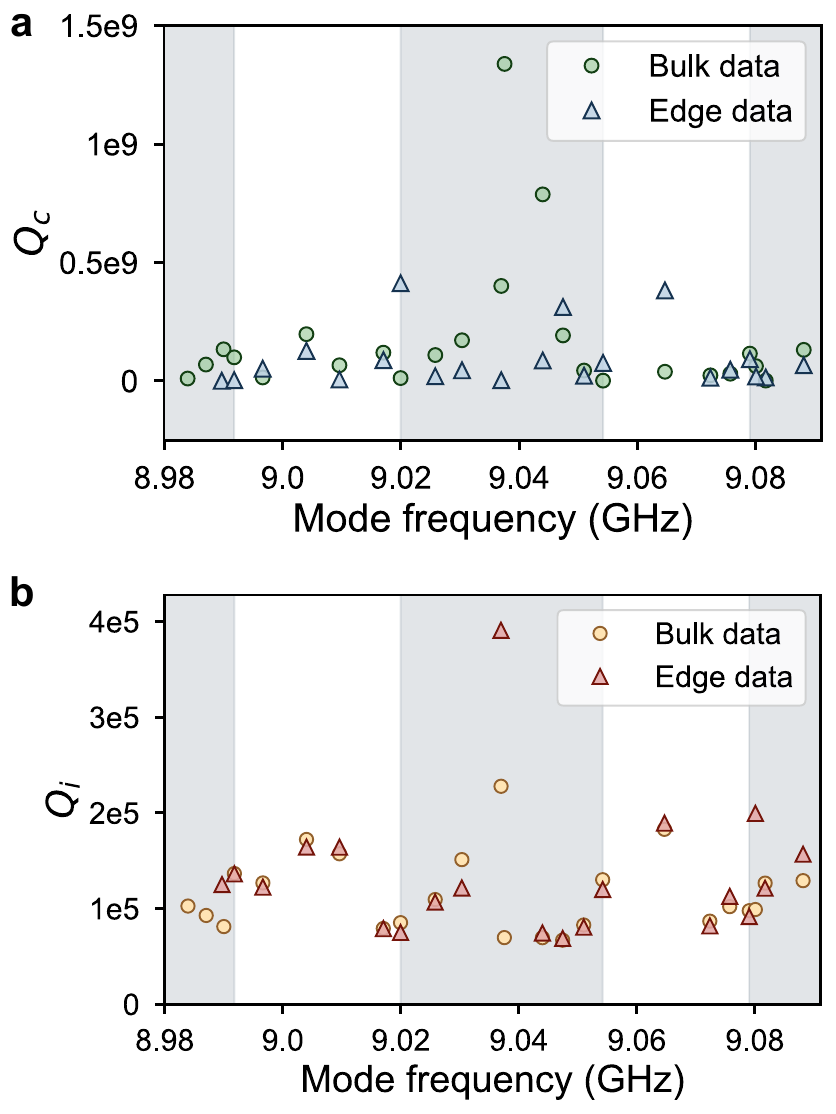}
    \internallinenumbersJ
    \caption{\textbf{Quality factors of lattice modes from edge and bulk spectra.} Internal ($Q_i$) and coupling ($Q_c$) quality factors are extracted from fits to modes in the lattice spectra displayed in Fig.~\ref{fig:SpectroscopyofLattice}b in the main text. Taken on the bare lattice prior to addition of the transmon qubit, these microwave transmission spectra were measured between a pair of lattice bulk sites (extracted Qs are represented by triangles in the plots) and between a pair of edge sites (extracted Qs are represented by circles in the plots). Expected locations of bulk bands, based on mode counting, are highlighted in gray and are intended to serve as a guide for the eye. Note that not all 25 lattice eigenmodes are apparent in the spectra and thus not all 25 eigenmodes are apparent in this plot. In \textbf{a,} coupling quality factors $Q_c$ range between $\approx$ 4.9e5 and 1.3e9. Towards the center of the band, some lattice eigenmodes with higher bulk participation fractions display relatively higher $Q_c$ values. In \textbf{b,} internal quality factors $Q_i$ range between  $6.7\times 10^4$ and $3.9\times 10^5$. In the gaps between gray bulk bands and in the very center of the spectrum, lattice eigenmodes with higher edge participation fractions display relatively higher $Q_i$ values, consistent with expectations that these modes suffer less loss due to their smaller relative participation in the lossier bulk sites equipped with YIG spheres. }
    \label{fig:CryoLatticeQsNoQubit} 
\end{figure}

\section{Lattice Disorder Characterization and Compensation}
Disorder in the lattice site frequencies must be controlled to a level below other energy scales of the system Hamiltonian (tunneling, particle-particle interactions, and magnetic field interactions). Lattice sites are tuned to degeneracy at room temperature, though the different types of lattice sites (single post, three post YIG-coupled) change frequency differently as the lattice sites cool to 20 mK. We first measure the change in frequency from cooling for the different lattice sites ($\sim23$ MHz for single post cavities and $\sim\!40$ MHz for cavities with YIG spheres) and then adjust for the differential at room temperature by adding indium to the top of the cavity posts, which effectively lengthens the cavity post and decreases the cavity frequency. Because indium is malleable and superconducting, it attaches easily to the cavity posts and does not decrease the cavity quality factors. After modifying post lengths with indium, we adjust each cavity frequency at 1K until the disorder in lattice site frequencies is less than $\pm 1$ MHz for the single post cavities and $\pm 3$ MHz for the YIG cavities. We tune the lattice to $\omega_l\approx 2\pi\times 8.9$ GHz for the measurements without a qubit.

\section{Probing Lattice Dynamics and Spectra}
\label{SI:LatticeDynamics}
In order to measure the linear response of the lattice, we insert dipole antennas into each site and connect them to a cryogenic switching network which is routed through circulators to enable performance of reflection measurements on most sites of the lattice (see SI~\ref{SI:DilFridgeWiring} for details on wiring and connectivity). To characterize the lattice, we first measure reflection on each site. Fig.~\ref{fig:SpectroscopyofLattice}d shows a pulse propagating on the edge of the lattice when the lattice edge site (1,1) is excited with a $80$ ns Gaussian pulse at the frequency of the green mode in Fig.~\ref{fig:SpectroscopyofLattice}c. The round trip time for an edge pulse is $\sim120$ ns compared to the decay time of $\sim 3\mu$s. The direction the pulse travels is set by the direction of the magnetic field and which band gap is excited. The two band gaps support edge channels of opposite chiralities.

\section{Transmon Characterization}
\label{SI:TransmonCharacterization}

The transmon chip is clamped in a copper holder that is then mounted on the side of the niobium lattice. Indium foil is added to the surface of the copper holder that clamps the qubit in order to better thermalize the sapphire chip on which the qubit is printed. There is one measurement port in the system strongly coupled to the readout cavity. All of the qubit measurements are performed via reflection measurements on the readout cavity.

The Hamiltonian of the 25 lattice sites and a readout resonator coupled to the qubit is the following:

\begin{equation}
\begin{split}
\label{eqn:ham1}
\frac{H}{\hbar} = \,& \sum_{p = q,r}\left(\omega_p \hat{a}_p^\dagger \hat{a}_p - \frac{\alpha_p}{2}\hat{a}_p^{\dagger 2} \hat{a}_p^2\right) - 2\chi_r \hat{a}_q^\dagger \hat{a}_q\hat{a}^\dagger_r \hat{a}_r\\
& + \sum_{l=1}^{N_{\text{modes}}=25} \left(\omega_l\hat{a}^\dagger_l \hat{a}_l - \frac{\alpha_l}{2} \hat{a}^{\dagger 2}_l \hat{a}_l^2 - 2\chi_l \hat{a}^\dagger_l \hat{a}_l \hat{a}^\dagger_q \hat{a}_q \right) \\
& + \sum_{l\neq m} -2\chi_{lm} \hat{a}^\dagger_l \hat{a}_l \hat{a}^\dagger_m \hat{a}_m,
\end{split}
\end{equation}

% \begin{equation}
% \begin{split}
% \label{eqn:ham1}
% \frac{H}{\hbar} = \,& \sum_{p = q,r}\left(\omega_p \hat{a}_p^\dagger \hat{a}_p + \frac{\alpha_p}{2}\hat{a}_p^{\dagger 2} \hat{a}_p^2\right) + \chi_r \hat{a}_q^\dagger \hat{a}_q\hat{a}^\dagger_r \hat{a}_r\\
% & + \sum_{l=1}^{N_{\text{modes}}=25} \left(\omega_l\hat{a}^\dagger_l \hat{a}_l + \frac{\alpha_l}{2} \hat{a}^{\dagger 2}_l \hat{a}_l^2 + \chi_l \hat{a}^\dagger_l \hat{a}_l \hat{a}^\dagger_q \hat{a}_q \right) \\
% & + \sum_{l\neq m} \chi_{lm} \hat{a}^\dagger_l \hat{a}_l \hat{a}^\dagger_m \hat{a}_m,
% \end{split}
% \end{equation}

where $\omega_q=\omega_{ge}$ is the $g\leftrightarrow e$ transition frequency of the transmon, $\alpha_q$ is the transmon anharmonicity (so $\omega_{ef}=\omega_q-\alpha$), $\omega_r$ is the bare readout resonator frequency, $\alpha_r$ is the self-Kerr shift of the readout resonator,  $2\chi_r$ is the qubit-readout dispersive shift, $\omega_l$ are the lattice mode frequencies, $2\chi_l$ are the qubit-lattice mode dispersive shifts, $\alpha_l$ are the self-Kerr shifts of the lattice modes, and $\chi_{lm}$ are the cross-Kerr shifts of the lattice modes. 

Initial qubit characterization was done with the single lattice site to which the qubit was coupled tuned to the lattice frequency ($\sim\!9$ GHz) while the other 24 sites were temporarily blocked off using screws lowered from the lid of the lattice until they contacted the post of the lattice site. This was done so that the simpler system of the qubit, readout resonator, and lattice resonator could be analyzed in isolation. Importantly, in this configuration the effects of the YIG-biasing magnets on the qubit lifetime could be directly measured by characterizing the qubit with and without the magnets present. This direct comparison would be difficult to make with every lattice site tuned to the lattice frequency since the magnets are required to achieve this tuning. We compare the measurements of the qubit with no magnets in the lattice to measurements in which the magnets were added to the lattice in Table~\ref{tab:MagnetContrast}. The introduction of the magnets resulted in the lifetime of the qubit dropping to $~3 \: \mu$s, approximately commensurate with the lifetimes of the cavity modes shown in Fig.~\ref{fig:SpectroscopyofLattice}. This suggests that improving the funneling of the magnetic field will be necessary to further increase the qubit lifetime.

\begin{table}
 \includegraphics[width=.48\textwidth]{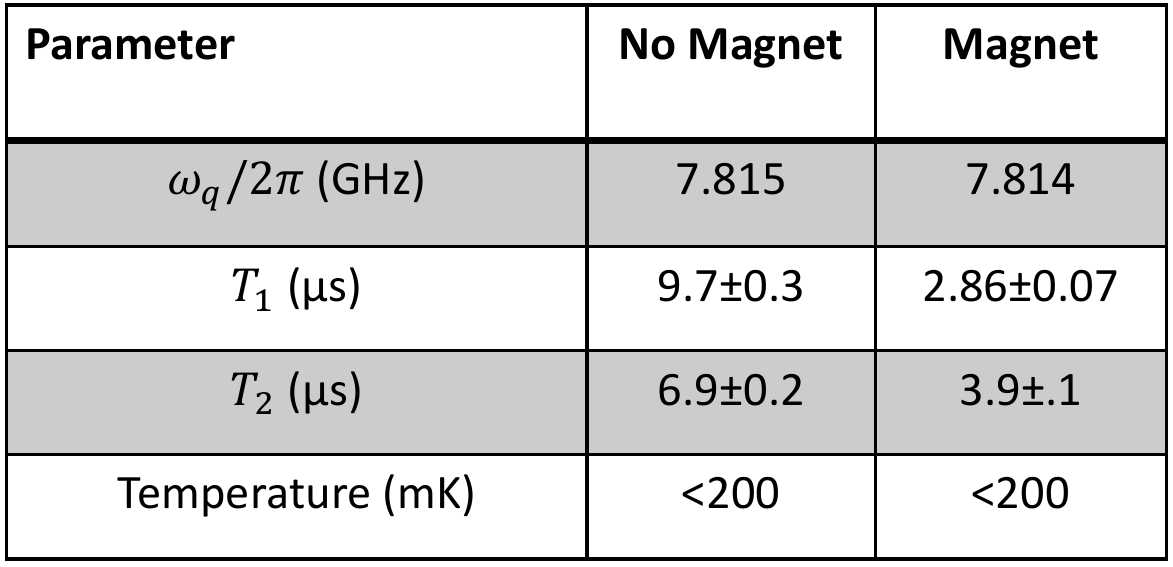}
 \internallinenumbersJ
 \caption{\textbf{Parameters of qubit coupled to a single lattice mode with and without a magnet in the lattice.} Adding the magnet decreases the lifetime of the qubit to $3\mu$s, a timescale similar to the lifetime of the cavity modes after the field is added.
 }
 \label{tab:MagnetContrast}
\end{table}

Next we characterized the qubit with every lattice site tuned to the lattice frequency and magnets added to the chiral site, so that the lattice was in the configuration described in Fig.~\ref{fig:SpectroscopyofLattice}, but with the qubit coupled to a single site on the corner of the lattice. Table~\ref{tab:qubit params} summarizes relevant parameters of this system and compares them to parameters found when only the single lattice site was tuned to the lattice frequency. When the lattice is tuned into resonance, the modes delocalize over the lattice, decreasing the electric field near the qubit antenna which in turn decreases the coupling to the qubit. To tune the lattice into resonance, the lattice cavities were temporarily decoupled and their bare frequencies individually tuned to $2\pi \times 8.901$ GHz. After the inter-site tunneling was restored in the lattice, the lattice modes' frequencies split into the chiral band structure described in Fig.~\ref{fig:SpectroscopyofLattice}. Qubit parameters were measured using standard techniques. The qubit frequency was measured through Ramsey spectroscopy. $\chi_r$ was measured by $\pi$-pulsing the qubit and measuring the shift in the readout resonator frequency. The $\chi_l$ shifts were measured by driving the cavity with a weak coherent tone and measuring the photon-number-dependent splitting in the qubit frequency. The self-Kerr $\alpha_l$ of the modes were calculated from the measured dispersive shifts $\chi_l$ and transmon anharmonicity $\alpha_q$, using the relation $2\chi_{l} = \sqrt{\alpha_q \alpha_l}$~\cite{Nigg_PRL2012}.

\begin{table}
 \includegraphics[width=.48\textwidth]{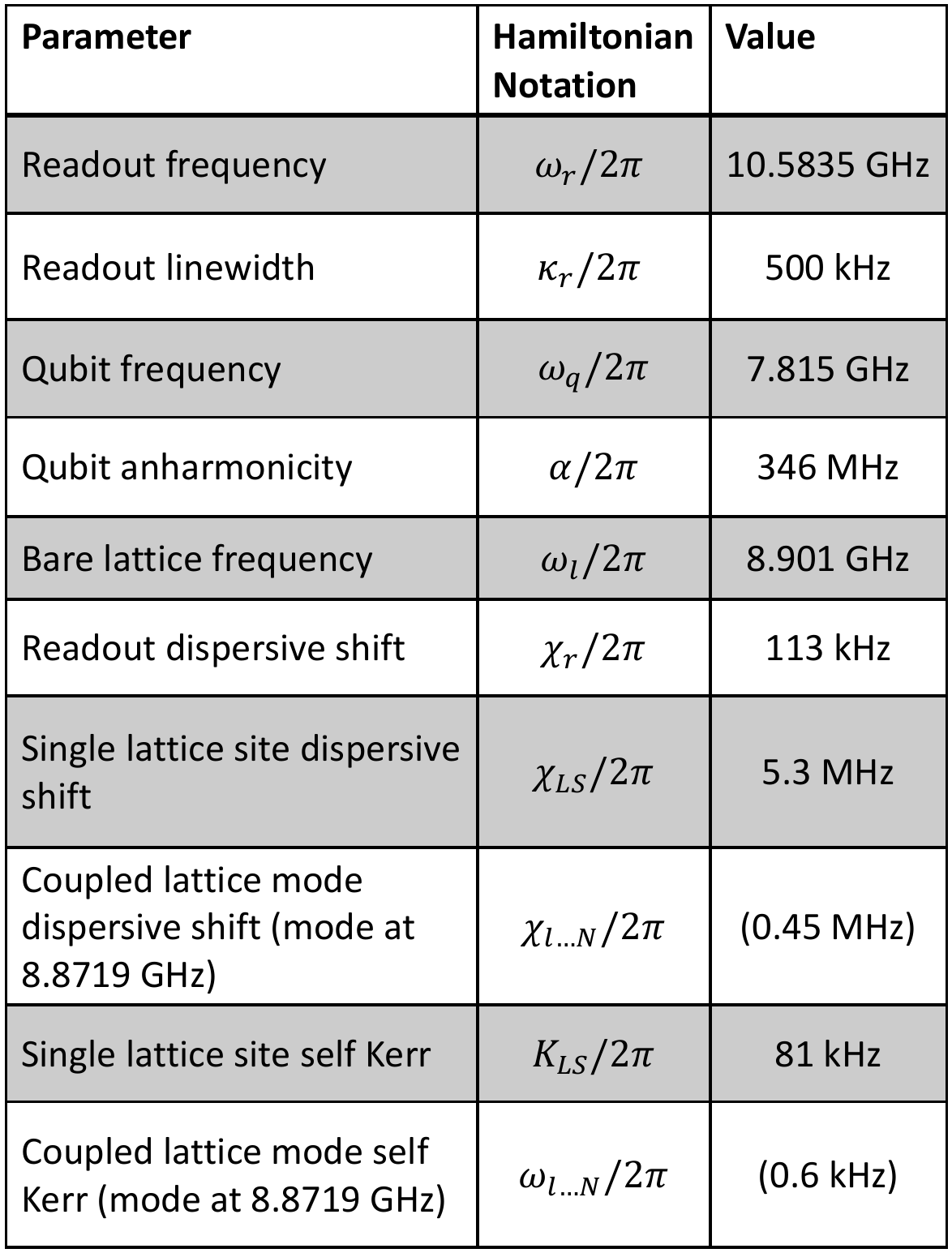}
 \internallinenumbersJ
 \caption{\textbf{Parameters of qubit/lattice/readout resonator system.} Terms designated ``single lattice site'' are measured when only the lattice site to which the qubit is directly coupled is tuned to the lattice frequency while the rest of the lattice sites are far detuned. The bare lattice frequency is the target frequency to which each lattice site is tuned before the couplings to adjacent lattice sites are enabled. Values in parentheses are example quantities for the lattice mode investigated in Fig.~\ref{fig:PhotonCounting}b and \ref{fig:PhotonCounting}c.
 }
 \label{tab:qubit params}
\end{table}

\section{Spectroscopy of the Lattice Modes Coupled to the Qubit}
\label{SI:LatticeSpec}

\begin{figure*}
 \centering
 %\begin{subfigure}[b]{\textwidth}
 \includegraphics[width=\textwidth]{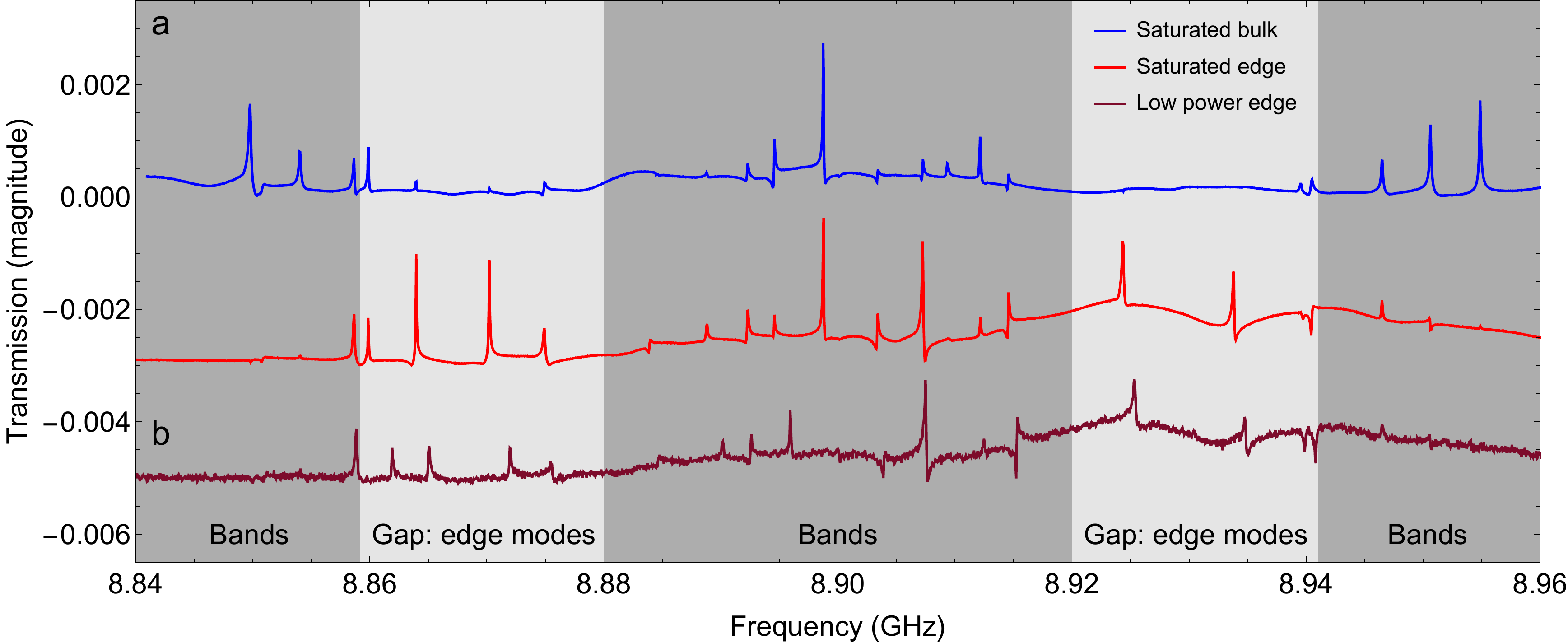}
 \internallinenumbersJ
 \caption{\textbf{Transmission spectroscopy through the lattice coupled to a qubit.} In \textbf{a,} the blue plot and the light red plot compare the transmission between two bulk cavities (site $(3,3)$ and $(2,3)$) and two edge cavities (sites $(1,4)$ and $(3,1)$) at high powers so that the qubit is saturated and the lattice is measured while effectively decoupled from the qubit. As predicted, two large gaps exist in the bulk-bulk transmission and modes in the edge-edge transmission reside in these gaps. As shown in Fig.~\protect{\SFigref{tab:latticemodes}}, there are four edge modes per band, but two of the edge modes are quite close to the bulk band and share some characteristics with the bulk modes. The dark red plot in \textbf{b} is the edge-edge transmission taken at lower power so that the modes are shifted by their coupling to the qubit. The modes shift in energy relative to the spectra at higher powers, showing that they are strongly coupled to the qubit.}
 \label{fig:LatticeWQubitCharacterization}
\end{figure*}

For these measurements, transmission is measured analogously to Fig.~\ref{fig:SpectroscopyofLattice}b, but with the qubit introduced into the lattice. Fig.~\SFigref{fig:LatticeWQubitCharacterization}a compares transmission between two edge sites (sites $(1,4)$ and $(3,1)$) and two bulk sites (site $(3,3)$ and $(2,3)$) at high powers, where the qubit is saturated and thus effectively decoupled from the lattice, causing the lattice modes to appear at their bare frequencies~\cite{reed2010high}. Similar to the spectra in the lattice without the qubit, the edge-edge transmission has modes located in the large gaps between the bulk modes. These measurements have a large background that was not present in the measurements in Fig.~\ref{fig:SpectroscopyofLattice}, suggesting that some direct coupling between the input and output lines was acquired while setting up the measurements with the qubit. 

\begin{figure*}
 \includegraphics[width=.98\textwidth]{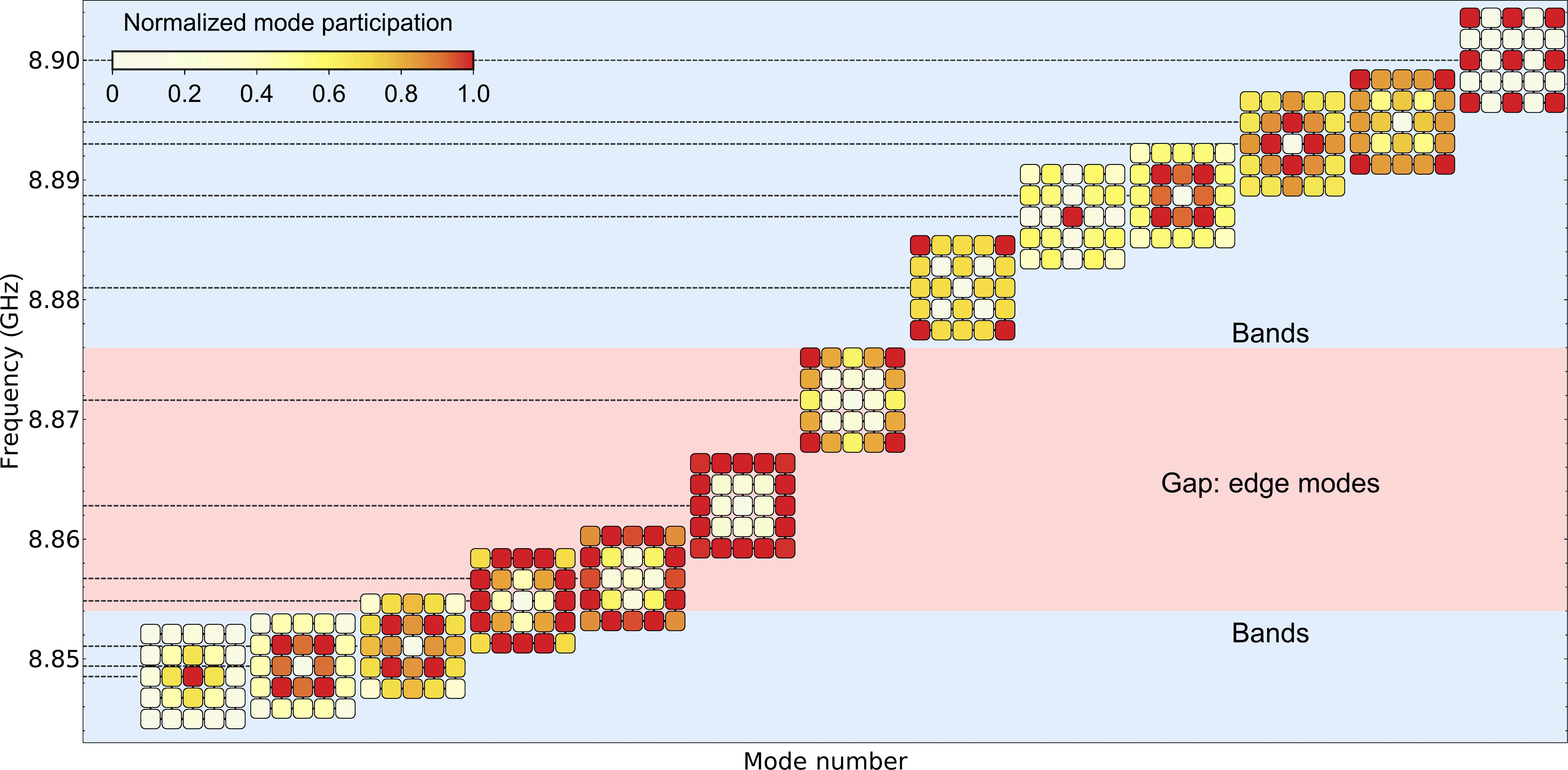}
 \internallinenumbersJ
 \caption{\textbf{Simulations of the eigenmodes of the $5\times 5$ quarter-flux Hofstadter Hamiltonian.} The eigenvectors of the first 13 modes of the lattice are calculated and plotted versus these modes' frequencies. Only these 13 modes are shown because they represent all unique eigenvectors of the system, as the modes are symmetric around the middle (13th) mode. In a system of this size, there are two modes that are localized strongly to the edge of the system, and two modes located in the gap but close enough to the bulk bands that they have more participation in the bulk lattice sites. The 13th mode is located where the two middle bulk bands touch and maintains its large delocalization for a lattice of any size. Each mode's color scale is normalized to the site in that mode with the largest participation.}
 \label{tab:latticemodes}
\end{figure*}

The lower frequency bulk band gap hosts three distinct edge modes between $8.86$ GHz and $8.88$ GHz, consistent with simulation of a $5\times 5$ quarter-flux Hofstadter model (see Fig.~\SFigref{tab:latticemodes}), while the higher frequency bulk band gap is seen to host only two distinct edge modes between $2\pi\times 8.92$ and $2\pi\times 8.94$ GHz. The third edge mode associated with this band gap is predicted to be much closer to the bulk modes as seen in Fig.~\SFigref{tab:latticemodes}. The top and bottom bulk bands are expected to have only four modes, so the peak doublet near $2\pi\times 8.94$ GHz is likely to be a hybridization of one bulk mode and one edge mode that are pushed closer in frequency than in the disorder-free model. This trend of the top band being compressed is consistent with the data taken in Fig.~\ref{fig:SpectroscopyofLattice} and is likely due to the presence of the opposite-chirality modes in the chiral cavities. These modes are located $2\pi\times 140- 2\pi\times 200$ MHz higher in frequency than the bare lattice frequency and are strongly coupled to the neighboring sites ($\sim\! 2\pi\times 20$ MHz). The presence of these modes can shift the higher frequency modes of the lattice by $\sim\!2\pi\times 4$ MHz and compress the upper band gap. Fig.~\SFigref{fig:LatticeWQubitCharacterization}b shows the same edge-edge transmission as probe power is varied. When the power is lowered, the lattice modes acquire a dispersive shift from the qubit that is proportional to the lattice corner site's participation in the mode.

Edge modes of the lattice experience power dependent shifts of $2\pi\times (1\sim 2)$ MHz while modes constrained to the bulk of the lattice shift by much less than the linewidth of the mode.

\setcounter{sfigure}{7}

We calculate the eigenmodes and eigenvectors of an ideal $5\times 5$ Hofstadter Hamiltonian (see Fig.~\SFigref{tab:latticemodes}), at a flux of $\frac{\pi}{2}$ per plaquette. We see excellent qualitative agreement of the eigenvalues of the modes with the data shown in Fig.~\ref{fig:SpectroscopyofLattice}. For a system this small there are only four modes located in each bulk band gap for a total of eight ``edge" modes. However, two of the edge modes in each gap are much closer to the bulk band frequency, which causes the eigenvectors of the these modes to have more participation in the bulk. In Fig.~\ref{fig:SpectroscopyofLattice}, these two edge modes that are isolated further from the edge mode have a stronger response in edge-edge transmission, while transmission through the modes located closer to the bulk bands starts to decrease in comparison. This transition from bulk to edge mode at the band edge is seen in larger systems as well. For future experiments where we will use the chiral channel to transport quantum states, photons must be transferred into a superposition of edge modes in order to create a localized traveling wave packet. These two modes located closest to the center of the band gap would be ideal modes to create a traveling single photon state, since they are isolated from the bulk and are thus more robust to disorder.

\setcounter{sfigure}{8}

\section{Structure of Pulsed Measurements with Qubit}
\label{SI:PulsedStructure}

\begin{figure}
    \centering
    \includegraphics[width=.48\textwidth]{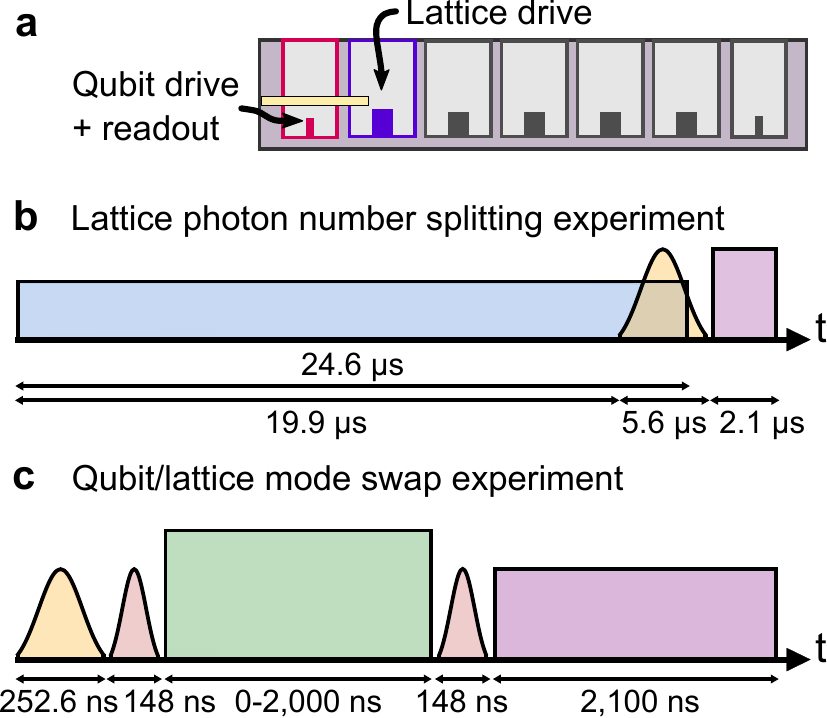}
    \internallinenumbersJ
    \caption{\textbf{Pulse sequences used for Fig.~\ref{fig:PhotonCounting}.}
    In \textbf{a,} Locations of qubit drive, readout tone, and drive on the lattice corner site are shown in a schematic cross-section of the lattice edge. The transmon chip is represented in yellow, while the readout cavity is boxed in red and the lattice corner site is boxed in blue. \textbf{b,} To measure the photon-number-dependent shift in the qubit resonance resulting from photons populating the lattice eigenmodes, we apply a very long (24.6 $\mu$s) weak drive (blue pulse) at lattice eigenmode frequencies to the corner site of the lattice which is directly coupled to the qubit. After 19.9 $\mu$s, we apply a weak Gaussian pulse of $\sigma = 1400$ ns (yellow) at $\omega_q$ and then read out (purple pulse) the qubit absorption after the lattice eigenmode drive is completed. \textbf{c,} To generate Rabi oscillations between qubit and a swath of the lattice eigenspectrum, we prepare the qubit in its $\ket{f}$ state with two $\pi$ pulses at $\omega_q = \omega_{ge}$ (yellow) and $\omega_{ef}$ (red). We drive dressed vacuum Rabi oscillations with a square pulse of varying length at $\omega_{f0\leftrightarrow g1} = (\omega_{ge} + \omega_{ef} - \omega_l)$, apply a $\pi$ pulse at $\omega_{ef}$ (red) to enable readout on the $|g\rangle \leftrightarrow |e\rangle$ transition, and read out (purple).}
    \label{fig:PulseChains}
\end{figure}

The transmon qubit is controlled and read out dispersively via drive and readout tones applied to its readout resonator, shown enclosed in the left side of the blue boxes in Fig.~\ref{fig:LatticeSetup}a and enclosed in red in Fig.~\SFigref{fig:PulseChains}a. To generate number splitting data like that shown in Fig.~\ref{fig:PhotonCounting}c, we populate relevant eigenmodes with photons by providing long and weak drive tones at these modes' frequencies through an antenna weakly coupled to the corner site of the lattice most proximate to, and directly coupled to, the qubit. While this weak drive is still being supplied to the lattice corner site, we excite the qubit from its ground state by supplying its readout resonator with a long Gaussian drive pulse ($\sigma = 1400$ ns) that has a fraction of the amplitude it would take to fully place the qubit in its first excited state. We then measure the degree to which the qubit is excited from its ground state with a readout tone applied at a range of frequencies in the vicinity of the original (zero-lattice-photon) qubit transition. In this way we can characterize the photon-number-dependent shift of the qubit resonance, which depends on photonic population in each lattice eigenmode that has enough edge participation to couple notably to the qubit. This pulse sequence is diagrammed in Fig.~\SFigref{fig:PulseChains}a. 

To generate Rabi oscillations between the qubit and lattice eigenmodes, as demonstrated in Fig.~\ref{fig:PhotonCounting}a and Fig.~\ref{fig:PhotonCounting}b, we supply a sequential set of $\pi$ pulses at the qubit transition frequencies $\omega_{ge}$ and $\omega_{ef}$ to prepare the qubit in its $\ket{f}$ state. We then supply a square pulse of varying length ($0-2000$ ns) to the qubit's readout resonator to tune the qubit into resonance with targeted lattice eigenmodes and drive oscillations between qubit and lattice. We finally supply an additional $\pi$ pulse at $\omega_{ef}$ before reading out the qubit state. This pulse sequence is diagrammed in Fig.~\SFigref{fig:PulseChains}b.

\setcounter{sfigure}{9}

\section{Mode Dependence of Dispersive Shift}
\label{SI:ModeDependenceShift}

The transmon qubit is directly coupled to a single corner site of the lattice, as depicted at top left of Fig.~\ref{fig:LatticeSetup}a and in Fig.~\ref{fig:LatticeSetup}c. If all other sites in the $5\times 5$ lattice are detuned from this resonator, each photon in the corner site induces a shift of the qubit $|g\rangle\leftrightarrow |e\rangle$ transition of $2\chi$, with $\chi_{LS}\approx \frac{ g_l^2}{\Delta}\times\frac{\alpha_q}{\Delta+\alpha_q}\approx 2\pi\times 5.3$ MHz. When all other lattice sites are tuned to resonance and hybridize forming the band structure, the dispersive shift is diluted between the modes of the band structure. Lattice mode $l$ experiences a dispersive shift $2 \chi_l=2\chi_{LS}\times |\langle u_l|LS\rangle|^2$, where $|u_l\rangle$ is the wavefunction of mode $l$, and $|LS\rangle$ is the wavefunction of a photon localized on the corner site coupled to the transmon. Note that $\sum_l \chi_l=\chi_{LS}$.

\begin{figure}
 \includegraphics[width=.48\textwidth]{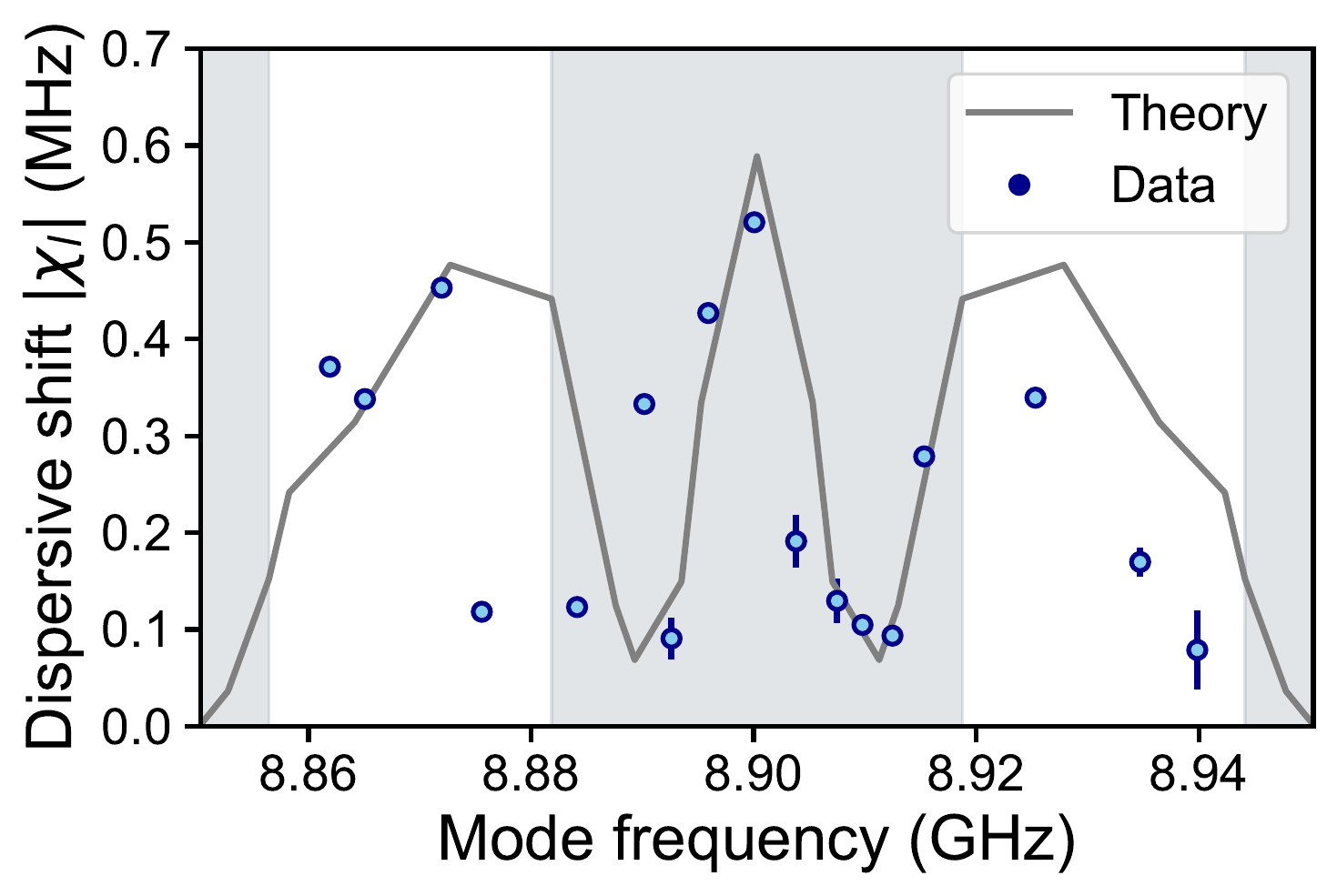}
 \internallinenumbersJ
 \caption{ \textbf{Mode-dependent dispersive shift of qubit.} Theoretical predictions of the dispersive shift between the qubit and each lattice eigenmode are compared to measured values from number splitting data. An example plot of number splitting due to population of a particular lattice eigenmode is seen in inset \textbf{c.} of Fig.~\ref{fig:PhotonCounting}. Theory is based on the measured (single-site) $\chi_{LS}$ scaled by the spectral weight of each lattice mode at the lattice corner site which is coupled to the qubit. Only modes present in the chevron spectrum of the lattice pictured in Fig.~\ref{fig:PhotonCounting} are included. The separation between 0- and 1-photon peaks when mode $l$ is driven is defined as $2 \chi_l$. Expected regions of the bulk bands, estimated from mode indices, are highlighted in gray. Error bars are derived from fits to number splitting data.}
 \label{fig:chisestimate}
\end{figure}

The mode employed in Fig.~\ref{fig:PhotonCounting}c, with index 7 in Fig.~\SFigref{tab:latticemodes}, exhibits a shift per photon of 2$\chi_{7}$ where $\chi_{7}= 2 \pi \times 0.45$ MHz, extracted from the frequency difference between zero- and one-photon resonances of Fig.~\ref{fig:PhotonCounting}c. This mode has a wavefunction overlap with the corner site of $|\langle u_7|LS\rangle|^2=0.09$, and thus we anticipate a shift of $2 \chi_{7}=2 \chi_{LS}\times|\langle u_{7}|LS\rangle|^2=2 \times 2\pi\times 5.3$ MHz $\times 0.09 \approx 2 \times 2\pi \times 480$ kHz, in agreement with the measured $\chi_{7}= 2 \pi \times 450$ kHz. In Fig.~\SFigref{fig:chisestimate} we compare, for each mode, the predicted shift with the shift extracted from the observed splittings between zero- and one-photon peaks of additional number splitting measurements.

\setcounter{sfigure}{10}

\section{Numerical Modeling of the \texorpdfstring{$|g0\rangle\leftrightarrow |f1\rangle$}{ } Oscillations in Fig.~\ref{fig:PhotonCounting}b}
\label{SI:3cnumerics}

To provide a fit for the stimulated vacuum-Rabi oscillations in the data plotted in Fig.~\ref{fig:PhotonCounting}b, we numerically simulate a model for a transmon qubit coupled to two modes through the $|g0\rangle\leftrightarrow |f1\rangle$ driven process.
We take into consideration the transmon population in the $\ket{g}$, $\ket{e}$, and $\ket{f}$ states along with the photon population in a selected pair of lattice eigenmodes over a timescale of 2 $\mu$s. We write the Hamiltonian of the qubit coupled to our selected subset of the lattice eigenspectrum in the rotating frame of the classical drive 

% \begin{equation}
% \begin{split}
% \label{eqn:ham2}
% H/\hbar = 
% \left( \omega_q - \omega_d \right)\hat{a}^\dagger_q \hat{a}_q + \frac{\alpha}{2} \hat{a}^\dagger_q \hat{a}^\dagger_q \hat{a}_q \hat{a}_q \\
% + \left(\omega_7 - \omega_d \right)\hat{a}^\dagger_7 \hat{a}_7 
% + g_7\left(\hat{a}^\dagger_q \hat{a}^\dagger_q \hat{a}_7 +  \hat{a}_q \hat{a}_q \hat{a}^\dagger_7 \right) \\
% + \left(\omega_8 - \omega_d\right)\hat{a}^\dagger_8 \hat{a}_8 
% + g_8 \left(\hat{a}^\dagger_q \hat{a}^\dagger_q \hat{a}_8 +  \hat{a}_q \hat{a}_q \hat{a}^\dagger_8 \right).
% \end{split}
% \end{equation}

% \begin{equation}
% \begin{split}
% \label{eqn:ham2}
% H/\hbar = 
% \left( \omega_q - \omega_d \right)\hat{a}^\dagger_q \hat{a}_q + \frac{\alpha_q}{2} \hat{a}^\dagger_q \hat{a}^\dagger_q \hat{a}_q \hat{a}_q \\
% + \sum_{l = 7,8}\left(\omega_l - \omega_d \right)\hat{a}^\dagger_l \hat{a}_l 
% + g_l\left(\hat{a}^\dagger_q \hat{a}^\dagger_q \hat{a}_l +  \hat{a}_q \hat{a}_q \hat{a}^\dagger_l \right).
% \end{split}
% \end{equation}

\begin{equation}
\begin{split}
\label{eqn:ham2}
H/\hbar = 
\left( \omega_q - \omega_d \right)\hat{a}^\dagger_q \hat{a}_q - \frac{\alpha_q}{2} \hat{a}^\dagger_q \hat{a}^\dagger_q \hat{a}_q \hat{a}_q \\
+ \sum_{l = 7,8}\left(\omega_l - \omega_d \right)\hat{a}^\dagger_l \hat{a}_l 
+ g_l\left(\hat{a}^\dagger_q \hat{a}^\dagger_q \hat{a}_l +  \hat{a}_q \hat{a}_q \hat{a}^\dagger_l \right).
\end{split}
\end{equation}

\begin{figure*}
    %\centering
    \includegraphics[width=.96\textwidth]{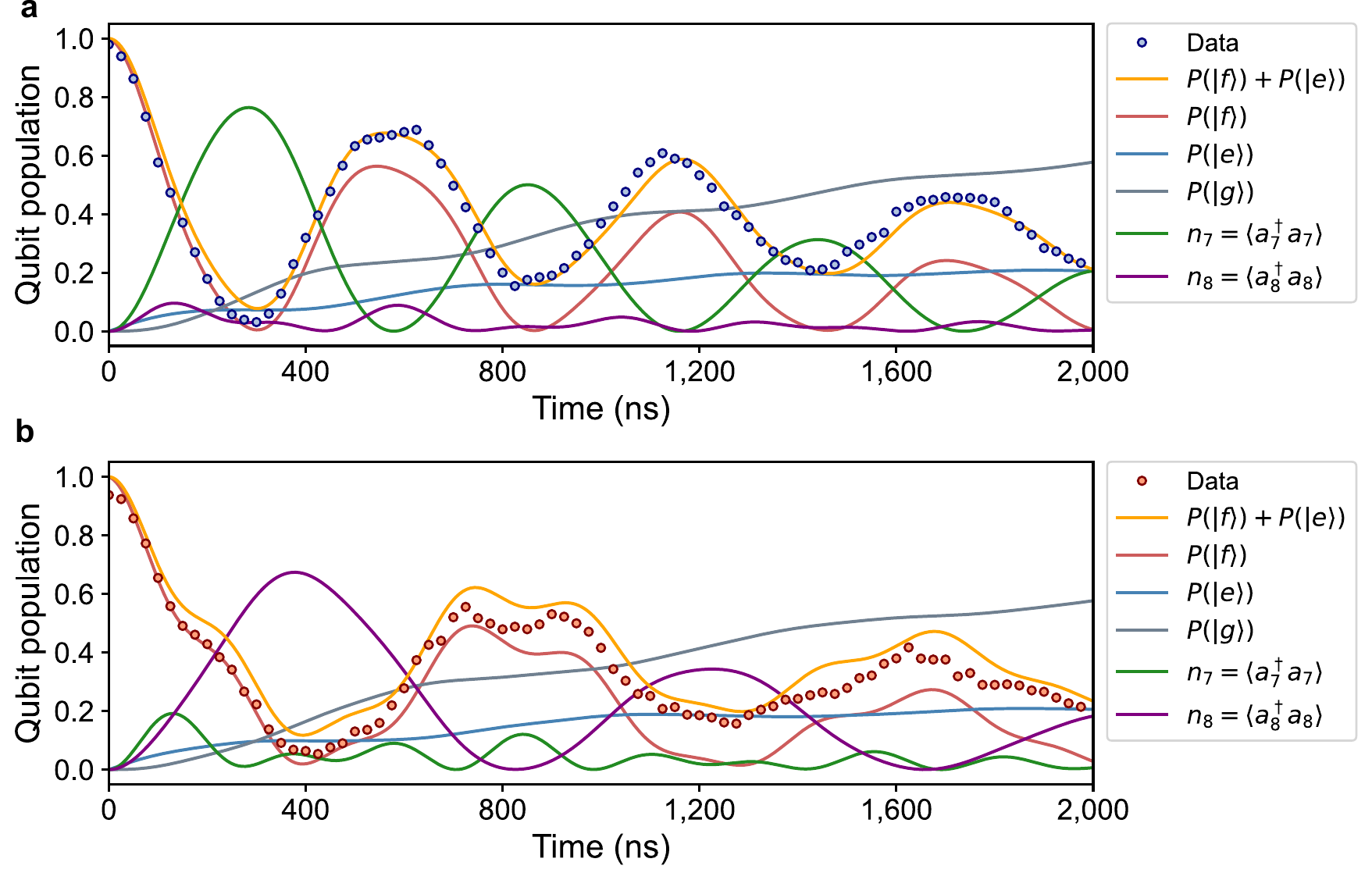}
    \internallinenumbersJ
    \caption{\textbf{Simulated qubit excited state population for Fig.~\ref{fig:PhotonCounting}b.}  Simulated populations $P(|f\rangle)$, $P(|e\rangle)$, and $P(|g\rangle)$ in the transmon qubit's $\ket{f}$, $\ket{e}$, and $\ket{g}$ states are plotted in time. In \textbf{a,} we retrieve a fit (yellow) to the measured data (blue points) plotted in Fig.~\ref{fig:PhotonCounting}b and taken at the frequency of the dressed $\ket{\tilde{e}}$ state singled out by blue line in Fig.~\ref{fig:PhotonCounting}a. To achieve this fit we represent the qubit excited state population as $P(|f\rangle)$ + $P(|e\rangle)$. Photonic occupancies $n_7$ (green) and $n_8$ (purple) of the eigenmodes selected for simulation are also plotted. It can be seen that the occupancy $n_7$ of the lattice eigenmode resonant with the dressed $\ket{\tilde{e}}$ state oscillates qualitatively out of phase with the qubit excited state population $P(|f\rangle)$ + $P(|e\rangle)$, consistent with a swap of excitation between qubit and mode. In \textbf{b,} we retrieve a fit (yellow) to the measured data (red points) taken at the frequency of the dressed $\ket{\tilde{e}}$ state close to resonance with a nearby lattice eigenmode to the one targeted in \textbf{a}. The frequency of the dressed $\ket{\tilde{e}}$ state at which this data was taken is singled out by a red line in Fig.~\SFigref{fig:numerics2d}a.}
    \label{fig:numerics1d}
\end{figure*}

\begin{figure*}
    \centering
    \includegraphics[width=.9\textwidth]{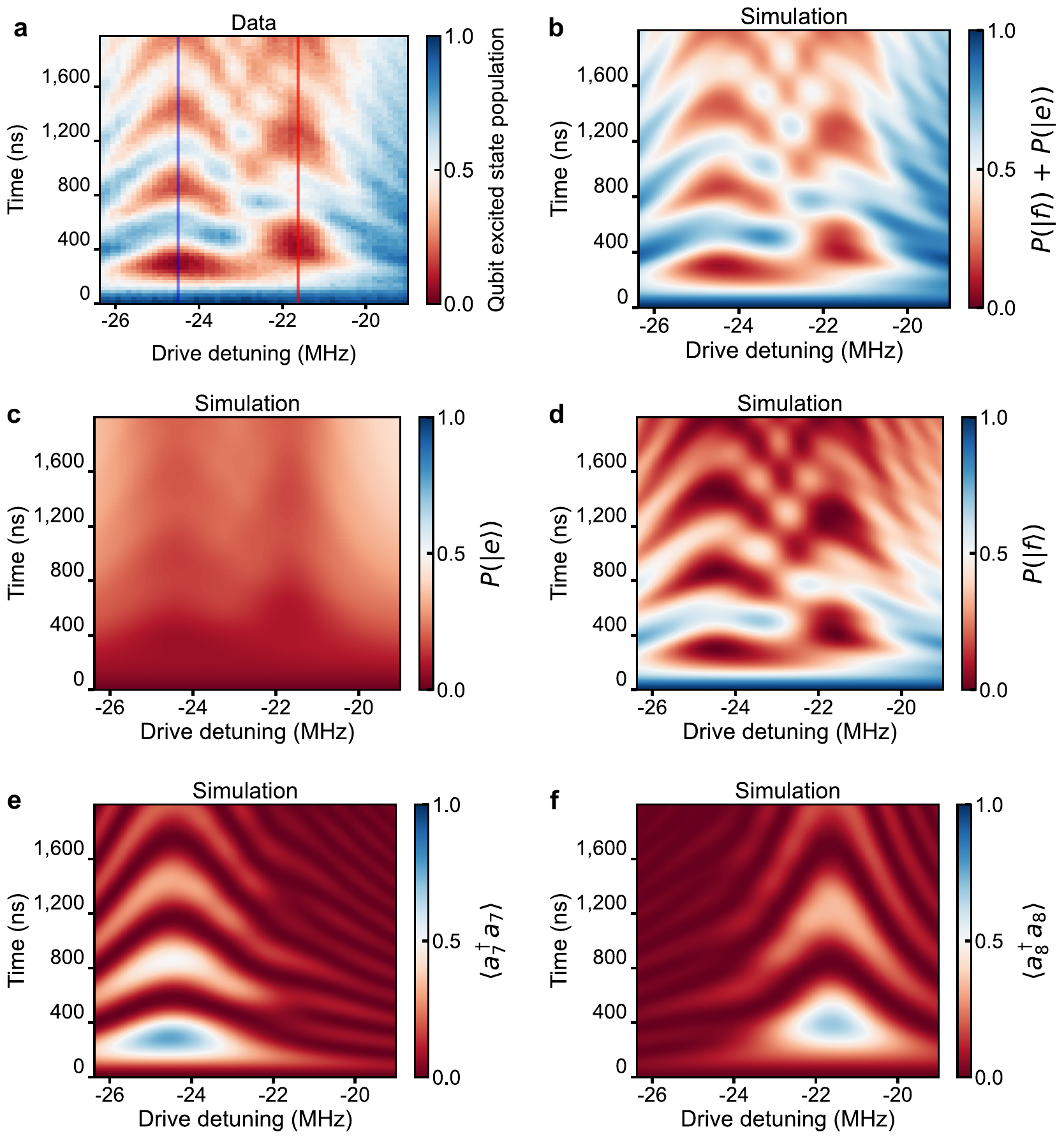}
    \internallinenumbersJ
    \caption{\textbf{Simulated observables for qubit/lattice eigenmode swap experiment} In \textbf{a,} a subset of Fig.~\ref{fig:PhotonCounting}a at a narrower range of dressed $\ket{\tilde{e}}$ state frequencies is overlaid with vertical lines which point out the frequencies at which data in Fig.~\SFigref{fig:numerics1d}a (blue line) and Fig.~\SFigref{fig:numerics1d}b (red line) were taken. \textbf{b,} Simulated qubit excited state population, rendered as $P(\ket{f}) + P(\ket{e})$, is plotted over a range of drive detunings from band center shared by the data in \textbf{a}. Note that this simulation is performed for a system of a qubit coupled to two lattice eigenmodes, and does not incorporate contributions from coupling to other regions of the lattice eigenspectrum. \textbf{c,} Simulated $P(\ket{e})$ is plotted for the range of drive detunings from band center used in \textbf{a}. \textbf{d,} Simulated $P(\ket{f})$ is plotted for the range of drive detunings from band center used in \textbf{a}. \textbf{e,} Simulated occupancy $n_7 = \langle \hat a_7^\dagger \hat a_7 \rangle$ of the seventh lattice eigenmode is plotted for the range of drive detunings from band center used in \textbf{a}. The drive frequency at which the qubit dressed $\ket{\tilde{e}}$ state is brought near resonance with this lattice eigenmode is overlaid with a blue vertical line in \textbf{a}. Note that this eigenmode is the edge mode for which swap data is plotted in Fig.~\ref{fig:PhotonCounting}b and for which $\chi_l = \chi_7$ and $\omega_l = \omega_7$ are quoted in Table~\ref{tab:qubit params}. 
    \textbf{f,} Simulated occupancy $n_8 = \langle \hat a_8^\dagger \hat a_8 \rangle$ of the eighth lattice eigenmode is plotted for the range of drive detunings from band center used in \textbf{a}. The drive frequency with which the qubit dressed $\ket{\tilde{e}}$ state is brought near resonance with this lattice eigenmode is overlaid with a red vertical line in \textbf{a}. Together, \textbf{e} and \textbf{f} illustrate that under the selected simulation parameters, it is possible to drive oscillations in the photonic occupancy of each of the pair of lattice eigenmodes separately by selecting different drive frequencies.}
    \label{fig:numerics2d}
\end{figure*}

The classical drive induces a resonant qubit-mode coupling at $\omega_d = 2\omega_q -\alpha_q - \omega_l$ for each mode $l \in \{7,8\}$. The dynamics is simulated by incorporating the photon loss in the transmon ($\Gamma$) and lattice modes ($\kappa$) and solving the master equation in the standard Lindblad form

\begin{equation}
    \label{eqn:mastereqn1}
    \dot{\rho} = - \frac{i}{\hbar}[H, \rho] + \mathcal{D}[\sqrt{\kappa}(\hat{a}_7 + \hat{a}_8)]\rho + \mathcal{D}[\sqrt{\Gamma}\hat{a}_q]\rho,
\end{equation}
using the usual definition for the damping superoperator $\mathcal{D}[\hat{L}] \rho = \hat{L} \rho \hat{L}^\dagger - \frac{1}{2}\{\hat{L}^\dagger\hat{L}, \rho\}$. In our numerics, we restrict the transmon and lattice modes to accommodate up to four excitations each and further truncate the Hilbert space to manifolds that have a maximum  of four total excitations in the combined system. Initializing the system with the qubit excited to its $\ket{f}$ state and the lattice modes empty of photons, we calculate the qubit excited state populations $P(|f\rangle)$ and $P(|e\rangle)$ during the exchange dynamics, along with photon populations $\hat{n}_l = \hat{a}^\dagger_l \hat{a}$ in the two lattice eigenmodes. The simulation results are displayed in Fig.~\SFigref{fig:numerics1d}. Further simulation results, plotted for a range of detunings of drive frequency from band center, are shown alongside data in Fig.~\SFigref{fig:numerics2d}. We find very good agreement with the data in Fig.~\ref{fig:PhotonCounting}b if the measured transmon excited state population includes population in the $|e\rangle$ state explained by the $|f\rangle \rightarrow |e\rangle$ decay during the dynamics.
Additionally, simulation results illustrate that it is possible to selectively drive exchange dynamics between the qubit and primarily a single lattice mode by selecting a drive so that the dressed state $\ket{\tilde{e}}$ is resonant with the targeted mode. 

\setcounter{sfigure}{11}

\setcounter{sfigure}{12}

\section{Estimating Uncertainties for Fig.~\ref{fig:PhotonCounting}}
\label{SI:estimatingerrors}
We estimate the errors for Fig.~\ref{fig:PhotonCounting}b \& c as the standard deviation of signal-free regions of the datasets. For Fig.~\ref{fig:PhotonCounting}b, the data comes from the slice at $t=0$ of Fig.~\ref{fig:PhotonCounting}a. For Fig.~\ref{fig:PhotonCounting}c, the data comes from the zero-drive-power slice of Fig.~\ref{fig:PhotonCounting}c, inset, for readout detunings $\delta\in [-5.5,-1.5]$ MHz.

\end{document}